\documentclass[11pt]{article}
\pdfoutput=1
\usepackage{jheppub}
\usepackage[T1]{fontenc}
\usepackage{color}
\usepackage{float}
\usepackage{url}
\usepackage{braket}
\usepackage{amsmath}
\usepackage{amsthm}
\usepackage{amssymb}
\usepackage{amsfonts}
\usepackage{graphicx}
\usepackage{caption}
\usepackage{subcaption}
\usepackage{multirow,array}
\usepackage{qcircuit}
\usepackage{hyperref}
\usepackage{comment}
\definecolor{red}{rgb}{1,0.,0}

\usepackage[mathscr]{eucal}
\usepackage[bbgreekl]{mathbbol}
\usepackage{mathtools}
\usepackage{tikz}
\usepackage[compat=1.1.0]{tikz-feynman}
\usepackage{feynmf}
\newcommand{\be}{\begin{equation}}
\newcommand{\ee}{\end{equation}}
\newcommand{\bea}{\begin{eqnarray}}
\newcommand{\eea}{\end{eqnarray}}

\def\({\left(}
\def\){\right)}
\def\[{\left[}
\def\]{\right]}

\usepackage{xcolor}
\colorlet{myPurple}{blue!40!red}

\usepackage{color}

\DeclareMathAlphabet{\pazocal}{OMS}{zplm}{m}{n}

\newcommand{\lag}{\pazocal{L}}

\subheader{\begin{flushright}
\texttt{\texttt{IFT-UAM/CSIC-24-120}}
\end{flushright}}

\title{Diving inside holographic metals}

\author[a]{Javier Carballo,}
\author[b]{Ayan K. Patra}
\author[b]{and Juan F. Pedraza}

\affiliation[a]{Mathematical Sciences and STAG Research Centre, University of Southampton, Highfield,\\ Southampton SO17 1BJ, UK}
\affiliation[b]{Instituto de F\'isica Te\'orica UAM/CSIC, Calle Nicol\'as Cabrera 13-15, Madrid 28049, Spain}

\emailAdd{j.carballo@soton.ac.uk}
\emailAdd{a.patra@csic.es}
\emailAdd{j.pedraza@csic.es}

\abstract{We investigate the gravitational dual of a fermionic field theory at finite temperature and charge density in two spatial dimensions, subject to a deformation by a relevant scalar operator. This makes a $(3+1)$-dimensional Einstein-Maxwell system coupled to a free fermion fluid, known as an electron cloud, undergo a holographic renormalization group flow. 
The inner (Cauchy) horizon is destroyed and the near-singularity metric instead adopts the form of a positive-$p_t$ Kasner cosmology, signaling the collapse of the Einstein-Rosen bridge. Previous studies have suggested that this collapse hinders direct probing of the singularity. Nonetheless, we propose and compute several CFT observables that characterize the interior and near-singularity geometries. These include the thermal $a$-function, which decays with a specific power of $p_t$ as nearly all CFT degrees of freedom are integrated out, and two-point correlators for neutral and charged operators, with the latter directly probing the singularity despite the positive-$p_t$. We also calculate characteristic velocities related to entanglement and complexity growth in the time-evolved thermofield double state, as well as the butterfly effect indicative of operator spreading. Notably, the deformed electron cloud features a Lifshitz IR fixed point and an additional Kasner trans-IR fixed point, absent in neutral RG flows.}

\begin{document}
\maketitle

\section{Introduction}

The study of strongly correlated electron systems is a cornerstone of modern condensed matter physics, revealing complex phenomena such as high-temperature superconductivity, quantum criticality, and non-Fermi liquid behavior. Traditional theoretical approaches often struggle to fully capture the rich dynamics of these systems. However, the advent of holographic duality has opened new avenues for understanding these phenomena \cite{Herzog:2009xv,McGreevy:2009xe,Horowitz:2010gk,Sachdev:2010ch,Sachdev:2010uj,Sachdev:2011wg,Iqbal:2011ae,Hartnoll:2016apf}. 

A particularly intriguing development is the concept of an \textit{electron star} (ES), a planar fluid of charged fermions in AdS space, held in equilibrium by gravitational and electromagnetic forces \cite{Hartnoll:2010gu,Hartnoll:2010ik}. As the ground state of a gravitating, charged ideal fluid at finite chemical potential, electron stars display low-energy emergent Lifshitz scaling characterized with a dynamical critical exponent $z$ and a smeared Fermi surface, indicative of quantum critical behavior. Consequently, electron stars serve as compelling models for investigating metallic quantum criticality, an area where traditional field theoretic methods face challenges due to the numerous gapless excitations at the Fermi surface that must be accounted for in a strongly interacting IR fixed point \cite{Hartnoll:2009ns}.

In \cite{Puletti:2010de}, holographic models of metals at finite temperature have been developed to extend the gravity dual description of electron stars. These models, known as \textit{electron clouds} (EC), describe a $(2+1)$-dimensional system of strongly interacting fermions at finite temperature and charge density. At zero temperature, the electron star represents a stable ground state, characterized by a gravitating, charged ideal fluid of fermions at a finite chemical potential. As temperature increases, the system transitions to an electron cloud configuration suspended over a charged black hole in AdS$_4$ space \cite{Puletti:2010de}. This setup features both an inner and an outer edge, balancing the black hole's gravitational pull with electrostatic repulsion. The fraction of the total charge carried by the electron cloud decreases with rising temperature, and at a critical temperature, the system undergoes a third-order phase transition. Beyond this critical temperature, the electron cloud vanishes, leaving only a charged black hole configuration. 

The phase transition between the electron cloud and the AdS$_4$-Reissner-Nordström (AdS-RN) black hole at the critical temperature $T_c$ is particularly intriguing due to its third-order nature. At finite temperature, this transition is marked by a continuous change in free energy density, smoothly interpolating between the low-temperature electron cloud phase and the high-temperature black brane phase. Examining the electrical conductivity during these transitions offers further insights. At finite temperature, the system's conductivity transitions smoothly from the high-temperature behavior characteristic of an AdS$_4$-RN black brane to the low-temperature behavior of an electron cloud, influenced by bulk fermions only within a finite radial band. This smooth transition underscores the continuous nature of the phase change and highlights the robustness of these models in describing strongly interacting fermion systems across varying temperature regimes. For more details on the physics of holographic electron stars and electron clouds see \cite{Hartnoll:2010xj,Puletti:2011pr,Medvedyeva:2013rpa,Allais:2013lha,Liu:2013yaa,Gran:2018jnt,DiNunno:2021eyf}.

In this paper, we aim to further investigate the dynamics of holographic metals at finite temperature by exploring the effects of a relevant deformation. Specifically, we will examine the Einstein-Maxwell-fluid system in $(3+1)$-dimensions coupled to a massive scalar field, to explore the critical behavior and transport properties of these systems. We dub the resulting geometries \textit{deformed electron clouds} (DEC). Adding a scalar field is known to significantly alter the interior structure of black holes, particularly near the singularity. For instance, in AdS-Schwarzschild black holes, a scalar field induces substantial backreaction, and leads to a Kasner-like singularity \cite{Frenkel:2020ysx} consistent with the standard BKL behavior \cite{Belinsky:1970ew,Belinskii:1973zz,Belinskii:1981vdw}.

Various deformations in other holographic models have revealed interesting interior dynamics, such as a rapid collapse or expansion of the Einstein-Rosen bridge, cosmological bounces, and Kasner inversions \cite{Hartnoll:2020rwq,Hartnoll:2020fhc,Sword:2021pfm,Sword:2022oyg,Wang:2020nkd,Mansoori:2021wxf,Liu:2021hap,Das:2021vjf,Hartnoll:2022rdv,Caceres:2022hei,Liu:2022rsy,Gao:2023zbd,DeClerck:2023fax,Oling:2024vmq,Cai:2023igv,Gao:2023rqc,Cai:2024ltu,Arean:2024pzo,Caceres:2024edr}. Within the AdS/CFT framework, these phenomena are reinterpreted as Renormalization Group (RG) flows in the dual CFT \cite{Caceres:2022smh}, with various CFT observables capturing certain information of the resulting bulk geometries \cite{Arean:2024pzo,Caceres:2024edr,Caceres:2022smh,Caceres:2021fuw,Bhattacharya:2021nqj,An:2022lvo,Auzzi:2022bfd,Hartnoll:2022snh,Mirjalali:2022wrg,Caceres:2023zhl,Blacker:2023ezy,Caceres:2023zft}. Motivated by these insights, we will examine here the metallic RG flows dual to the DEC gemetries, focusing on their interior and near-singularity structure as well as specific CFT imprints.

This paper is structured as follows. Section \ref{sec2} provides a detailed review of the electron star and electron cloud models, highlighting their implications for transport properties at finite temperature. In Section \ref{sec3}, we introduce a relevant deformation to this system and examine its impact on the near-horizon and internal dynamics of black holes. Section \ref{sec4} focuses on various probes that can capture the effects of this deformation and can be used to extract information about the interior geometry. Finally, Section \ref{sec:discussion} summarizes our key findings regarding the bulk dynamics and discusses their implications for the dual CFT.

\section{The electron cloud system: a simple model for holographic metals\label{sec2}}

In this section, we will give a quick overview of the so-called electron star and electron cloud geometries~\cite{Hartnoll:2010gu,Hartnoll:2010ik,Puletti:2010de,DiNunno:2021eyf}. These geometries are argued to be holographic duals of $(2+1)$-dimensional strongly interacting fermions at finite temperature and charge density, commonly referred to as \emph{holographic metals}. These systems are described in the bulk as an Einstein-Maxwell theory coupled to fermions, where fermions are treated locally as an ideal fluid of charged free particles. Such a description is largely inspired by the standard Tolman-Oppenheimer-Volkoff treatment of neutron stars in general relativity \cite{Tolman:1939jz,Oppenheimer:1939ne}, hence the term \emph{electron stars}.  \emph{Electron clouds}, on the other hand, refer to their finite temperature generalizations. We will work with the latter and often use the term `electron cloud' to refer to the fermion fluid component of the system alone.

\subsection{Bulk action and equations of motion}

Besides the gravitational field, the bulk system includes a Maxwell field and an ideal fluid of charged fermions. Precisely, the action of the system we consider is
\be
	\label{eq:EC_action}
	S = S_{\text{EH}} + S_{\text{M}} + S_{\text{F}} = \int d^4x \sqrt{-g}(\pazocal{L}_{\text{EH}} + \pazocal{L}_{\text{M}} + \pazocal{L}_{\text{F}}) \; ,
\ee
where the Einstein-Hilbert, Maxwell, and charged perfect fluid Lagrangian densities are
\bea
	\lag_{\text{EH}} &=& \frac{1}{2 \kappa^2}(R - 2\Lambda) \;, \label{eq:lag_EH}\\
	\lag_\text{M} &=& -\frac{1}{4e^2}F_{\mu\nu}F^{\mu\nu} \;, \label{eq:lag_Fsquared}\\
	\lag_{\text{F}} &=& -\rho(\sigma)+\sigma u^{\mu}(\partial_{\mu}\Phi+A_{\mu})		+\lambda(u^{\mu}u_{\mu}+1) \;, \label{eq:lag_fluid}
\eea
respectively. Here $R$ is the Ricci scalar, $\Lambda$ is the cosmological constant,\footnote{We will work in units where the AdS radius is $L=1$. This amounts to setting $\Lambda=-3$.} $\kappa^2=8\pi G_N$, $F = dA$ is the electromagnetic field strength, and
$\rho$, $\sigma$ and $u^{\mu}$ are the fluid's energy density, charge density, and proper velocity respectively. Further, the fields $\Phi$ and $\lambda$ in $\lag_{\text{F}}$ are Lagrange multipliers that we must set on-shell. Variation with respect to these variables yields
\be
\nabla_{\mu}J^{\mu} \equiv \nabla_{\mu}(\sigma u^{\mu}) = 0 \;, \label{eq_ch4_1_Jfluid}
\ee
and
\be
u^2 = -1 \;,\label{eq:ch4_1_u2}
\ee
respectively. Equation \eqref{eq_ch4_1_Jfluid} defines a conserved fluid current vector, $J^{\mu}=\sigma u^{\mu}$, while \eqref{eq:ch4_1_u2} indicates that the fluid's velocity must be timelike. Also, as it is manifest, $\Phi$ ensures invariance under gauge transformations $A_{\mu} \rightarrow A_{\mu} + \partial_{\mu}\Phi$.

Varying the action with respect to $\sigma$ and $u^{\mu}$, one arrives at
\be
    \mu \equiv \rho'(\sigma) = u^{\mu}(\partial_{\mu}\Phi+A_{\mu}) \;,\label{eq:ch4_1_mu}
\ee
\be
\sigma(\partial_{\mu}\Phi + A_{\mu}) + 2\lambda u_{\mu} = 0 \;.
\ee
The first equation defines $\mu \equiv \rho'(\sigma)$, which we can interpret as a local chemical potential.
Introducing the perfect fluid pressure, $p= -\rho + \sigma \mu$, the second equation gives
\be
	\lambda = \frac{\sigma \mu}{2}=\frac{\rho + p}{2} \;.
\ee

Now that we have worked out the fluid section, we can move on to the equations of motion for the bulk fields. These can be obtained by varying the action with respect to $A_\mu$ and $g_{\mu\nu}$. As standard, the final result can be written as an Einstein-Maxwell system with sources,
\be
	\label{eq:EC_eom_gen}
	G_{\mu\nu} - 3g_{\mu\nu} = \kappa^2 (T_{\mu\nu}^{\text{M}} + T_{\mu\nu}^{\text{F}})\;, \quad \nabla^{\nu}F_{\mu\nu} = e^2 J_{\mu}\;,
\ee
where the stress-energy tensors and current density are given by
\bea
	T_{\mu\nu}^{\text{M}} &=& \frac{1}{e^2}\left ( F_{\mu}^{\phantom{\sigma}\sigma}F_{\nu\sigma} - \tfrac{1}{4} g_{\mu \nu} F_{\alpha \beta}F^{\alpha \beta}  \right )\;, \label{eq:strener_M}\\
	T_{\mu\nu}^{\text{F}} &=& (\rho + p)u_{\mu}u_{\nu} + pg_{\mu\nu}\;, \label{eq:strener_F}\\
	J_{\mu} &=& \sigma u_{\mu} \;, \label{eq:currentdens}
\eea
respectively.

\subsubsection{Details on the fermion fluid thermodynamics\label{sec:FFT}}
Before we attempt to solve these equations, there are some aspects of the fluid's thermodynamics we need to revisit closely following \cite{Hartnoll:2010gu}.

We consider a perfect fluid as an effective description of a system of charged fermions with mass $m$ and at zero temperature, following the ideal Fermi gas model.  One important point is that the ideal Fermi gas model does not inherently account for particle interactions, which is a critical aspect of strongly interacting systems. 
Since the applicability of this model as the holographic counterpart for strongly interacting fermions may seem questionable, it is essential we address this concern. First, we point out that after some dimensional analysis and simple estimations, one can show that gravitational and electromagnetic interactions between bulk fermions are locally negligible \cite{Hartnoll:2010gu}. Furthermore, we recall that known instances of AdS/CFT frequently involve generalized free fields in the bulk, which effectively map to strongly interacting QFTs in the boundary. 

Bearing this in mind, we define the energy and charge densities as
\be
    \label{eq:ch4_2_rhosigma}
    \rho = \int^{\mu}_{m}E\,g(E)dE\;, \quad \sigma = \int^{\mu}_{m}g(E)dE \;,
\ee
where $E$ is the single-particle energy, $\mu$ the chemical potential and $m$ the mass of the fermions. We set the fermionic electric charge to be one in units according to~\eqref{eq:lag_Fsquared}. The density of states, $g(E)$, for this system is known to be
\be
    \label{eq:ch4_2_densstates}
    g(E) = \beta E \sqrt{E^2 - m^2} \;,
\ee
where $\beta$ is just the proper dimensionful constant prefactor. Additionally, the equation of state for this system follows from the first law of thermodynamics in the grand canonical ensemble at zero temperature\footnote{Note that, for our choice of fermionic electric charge, number density and charge density are equal.}
\be
    \label{eq:ch4_2_eqstate}
    p = -\rho + \mu \sigma \;.
\ee

Let us now work with dimensionless quantities, which we denote with a `hat' symbol. From the equations of motion, we can infer
\be
    \label{eq:ch4_2_dimenless_1}
    A = \frac{e}{\kappa}\hat{A}\;, \quad \sigma = \frac{1}{e\kappa}\hat{\sigma}\;,\quad \rho = \frac{1}{\kappa^2}\hat{\rho}\;,\quad p = \frac{1}{\kappa^2}\hat{p}\;,
\ee
which, together with~\eqref{eq:ch4_2_eqstate}, imply
\be
    \label{eq:ch4_2_dimenless_2}
    \mu = \frac{e}{\kappa}\hat{\mu} \;.
\ee
Thus, the dimensionless energy density, charge density and pressure now read
\be
    \label{eq:ch4_2_rhosigma_hat}
    \hat{\rho} = \hat{\beta} \int^{\hat{\mu}}_{\hat{m}}\hat{E}^2\sqrt{\hat{E}^2-\hat{m}^2}d\hat{E} \;, \quad \; \hat{\sigma} = \hat{\beta} \int^{\hat{\mu}}_{\hat{m}}\hat{E}\sqrt{\hat{E}^2-\hat{m}^2}d\hat{E} \;, \quad \; \hat{p} = -\hat{\rho} + \hat{\mu}\hat{\sigma}\;,
\ee
respectively. Note that there are two unspecified parameters $\hat{\beta}$ and $\hat{m}$. From considerations described in~\cite{Hartnoll:2010gu}, one can show that our approximations are valid when
\be
    \label{eq:ch4_2_betam_hat}
    \hat{\beta} \sim 1 \;, \quad \; \hat{m}^2 \sim 1 \;.
\ee

\subsection{Exterior solutions and physical properties}
\label{s2_2}

The bulk geometry of an electron cloud system is characterized by a ($3+1$)-dimensional planar AdS-RN black hole coupled to the charged fermion fluid.

The ansatz we use to solve the geometry of this system in Poincaré coordinates is
\be
    \label{eq:EC_metric_ansatz}
    ds^2 = \frac{1}{v^2}\left ( -f(v)e^{-\chi(v)} dt^2 + \frac{dv^2}{f(v)} + dx^2 + dy^2 \right ) \;, \quad A = \frac{e}{\kappa}\hat{A} = \frac{e}{\kappa} h(v)dt \;,
\ee
where $v$ is the holographic radial coordinate; $v \rightarrow 0$ corresponds to the AdS boundary and $v \rightarrow \infty$ to the singularity. Note that all the metric and gauge functions are only dependent on $v$, as a consequence of staticity and isotropy in the spatial boundary coordinates. With this ansatz, Maxwell's equations~\eqref{eq:EC_eom_gen} imply that the only non-zero component of the fluid's velocity is $u^t$, i.e. the fermion fluid is also static. Therefore, its thermodynamic variables are also radial functions $p=p(v)$, $\rho=\rho(v)$, $\sigma = \sigma(v)$.

For convenience in our subsequent exploration of the interior of this black hole under a scalar deformation, we change to ingoing Eddington-Finkelstein (IEF) coordinates $u = t + v_*$, where $dv_* = \frac{e^{\chi(v)/2}}{f(v)}dv$, such that
\bea
    \label{eq:ch4_3_ansatz_EF}
        ds^2 &=& \frac{1}{v^2}\left ( -f(v)e^{-\chi(v)} du^2 + 2e^{-\frac{\chi(v)}{2}}dudv + dx^2 + dy^2 \right ) \;, \\ 
        \hat{A} &=& h(v)du - \frac{e^{\chi(v)/2}h(v)}{f(v)}dv \;.
\eea
To be consistent with the considerations above, we take $u^u$ to be the only non-zero component of the fluid's proper velocity. The radius of the outer event horizon, $v=v_+$, is a free parameter of the system due to the dilatation symmetry $v\rightarrow\lambda v$ of AdS. We make use of this symmetry
\be
    \label{eq:ch4_3_scaling}
    x^{\mu} \, \rightarrow \, \hat{x}^{\mu} \equiv \lambda x^{\mu} \;, \quad v_+ \, \rightarrow \, \hat{v}_+ \equiv \lambda v_+ \;, \quad h(v) \, \rightarrow \, \hat{h}(v) \equiv \lambda^{-1} h(v) \;,
\ee
to fix the horizon radius to one $\hat{v}_+=1$ by choosing $\lambda = v_{+}^{-1}$.

Regarding the chemical potential, we define it locally in order to determine the thermodynamics of the fluid in the bulk. By the AdS/CFT dictionary~\cite{Natsuume:2014sfa}, we can read off the chemical potential of the dual field theory from the gauge field's temporal component near-boundary expansion as its leading order term
\be
    \label{eq:ch4_3_nearbdy_A}
    \hat{A}_t\rvert_{\hat{v}=0} = \hat{\mu} \;.
\ee
However, a physical quantity must be gauge invariant, so one takes $\hat{A}_t\rvert_{\hat{v}=\hat{v}_+}=0$ such that the gauge invariant definition
\be
    \label{eq:ch4_3_mugaugeinv}
    \hat{\mu} = \hat{A}_t\rvert_{\hat{v}=0} - \hat{A}_t\rvert_{\hat{v}=\hat{v}_+} 
\ee
is equivalent to~\eqref{eq:ch4_3_nearbdy_A}. 

Finally, we define the bulk chemical potential $\hat{\mu}_{loc}$\footnote{We omit the `loc' subscript in the rest of the text for ease of writing.} to be
\be
    \label{eq:ch4_3_muloc_def}
    \hat{\mu}_{loc}(\hat{v}) \equiv  u^{\mu}(\hat{v})\hat{A}_{\mu}(\hat{v}) = u^t(\hat{v})\hat{A}_t(\hat{v}) = u^u(\hat{v})\hat{A}_u(\hat{v}) = \frac{e^{\chi(\hat{v})/2}\hat{v}}{\sqrt{f(\hat{v})}} \hat{h}(\hat{v}) \;,
\ee
where in the second equality we change to IEF coordinates and, in the last equality, we use~\eqref{eq:ch4_1_u2} to determine $u^u(\hat{v})$.

Using the equation of state~\eqref{eq:ch4_2_rhosigma_hat}, we arrive at the following independent equations of motion
\bea
    \hat{h}'' + \frac{1}{2}\chi'\,\hat{h}' - \frac{\hat{\sigma}\,e^{-\chi/2}}{\hat{v}^3\sqrt{f}} &=& 0 \;, \label{eq:ch4_3_eom_1}\\
    \hat{v}^4 e^{\chi} \,\hat{h}'^2 - 2\hat{v}f' + 6f + 2\hat{\rho} - 6 &=& 0 \;, \label{eq:ch4_3_eom_2}\\
    \frac{\chi'}{\hat{v}} - \frac{\hat{h}\,\hat{\sigma}\,e^{\chi/2}}{\hat{v}f^{3/2}} &=& 0 \;, \label{eq:ch4_3_eom_3}
\eea
where $'$ denotes $\frac{d}{d\hat{v}}$. The fluid variables, $\hat{\sigma}(\hat{v})$ and $\hat{\rho}(\hat{v})$, are determined analytically via~\eqref{eq:ch4_2_rhosigma_hat} giving
\bea
    \hat{\sigma}(\hat{v}) &=& \frac{\hat{\beta}}{3} \left ( \hat{\mu}(\hat{v})^2 - \hat{m}^2 \right )^{3/2} \;, \label{eq:ch4_3_fluid_sigmahat}\\
    \hat{\rho}(\hat{v}) &=& \frac{\hat{\beta}}{8} \left [ \hat{m}^4 \, \log \left ( \frac{\hat{m}}{\hat{\mu}(\hat{v}) + \sqrt{\hat{\mu}(\hat{v})^2 - \hat{m}^2}} \right ) + \hat{\mu}(\hat{v})\sqrt{\hat{\mu}(\hat{v})^2 - \hat{m}^2}\, (2\hat{\mu}^2 - \hat{m}^2) \right ] \;, \label{eq:ch4_3_fluid_rhohat}
\eea
where $\hat{\mu}(\hat{v})$ is the local chemical potential in~\eqref{eq:ch4_3_muloc_def}.

We are then left with three functions to solve for, $f(\hat{v}),\,\chi(\hat{v}),\,\text{and}\;\hat{h}(\hat{v})$, and three coupled ODEs. In order to numerically solve this system in the exterior of the black hole, we integrate the equations of motion from $\hat{v} = 1 - \epsilon$ (for $\epsilon \ll 1$), since we are fixing the horizon at $\hat{v}_+=1$, to $\hat{v}=\epsilon$, where the boundary resides. During this integration, we search for the region in which the fermion fluid has support. The condition to be met for the fluid to exist is easily derived from~\eqref{eq:ch4_2_rhosigma_hat} (notice the limits of integration) to be 
\be
    \label{eq:ch4_3_fluidcond}
    \hat{\mu}^2 > \hat{m}^2 \;.
\ee
Thus, we obtain the locations of the edges of the electron cloud from \eqref{eq:ch4_3_fluidcond} as we integrate the equations of motion.

Regarding the initial conditions, $f$ vanishes at the horizon, $f(\hat{v}_+)=0$, and so does the gauge potential, $\hat{h}(\hat{v}_+)=0$. For the rest of the conditions, we argue there must always be a gap with no fermion fluid between the electron cloud and the horizon, since $\hat{h}(\hat{v}_+)=0$ and $\hat{\mu}(\hat{v}) \propto \hat{h}(\hat{v})$. Therefore, the geometry near the horizon must tend to pure AdS$_4$-RN, characterized by\footnote{This can be verified by setting the fluid to zero ($\hat{\rho}=\hat{\sigma}=0$) in the equations of motion above and checking that the functions below satisfy the resulting equations.}
\bea
    f_{\text{\tiny{RN}}}(\hat{v}) &=& 1 + \frac{\hat{q}^2}{2}\hat{v}^4 - \left ( 1 + \frac{\hat{q}^2}{2} \right )\hat{v}^3 \;, \label{eq:ch4_3_RNfuncts_f} \\
    \chi_{\text{\tiny{RN}}}(\hat{v}) &=& 0 \;, \label{eq:ch4_3_RNfuncts_chi}\\
    \hat{h}_{\text{\tiny{RN}}}(\hat{v}) &=& \hat{q}(1-\hat{v}) \;, \label{eq:ch4_3_RNfuncts_h}
\eea
where $\hat{q}$ is the dimensionless black hole's electric charge.

\begin{figure}[t]
    \centering
    \includegraphics[width=\textwidth]{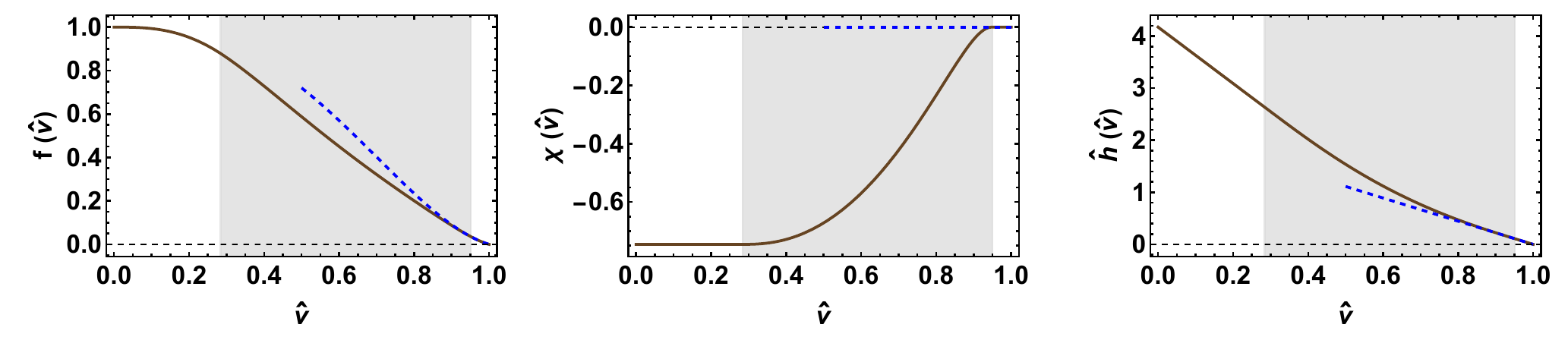}
    \caption{Numerical solutions to the metric functions, $f(\hat{v})$ (left) and $\chi(\hat{v})$ (middle), and gauge field, $\hat{h}(\hat{v})$ (right), from the boundary ($\hat{v} \rightarrow 0$) to the outer event horizon ($\hat{v}_+ = 1$) in the EC system. The region where the electron cloud is located is shown by the shaded gray area, and the blue-dashed line denotes the AdS$_4$-RN solution. This solution corresponds to $T/\hat{\mu} \approx 0.01$ and values for the input parameters: $\hat{\beta}=10$, $\hat{m}=0.55$ and $\hat{q} \simeq 2.23\,$.}
    \label{fig:ch4_3_estar_metricfuncts_all}
\end{figure}

The temperature of the black hole corresponding to \eqref{eq:EC_metric_ansatz} is given by
\be
    \label{eq:kasner_BH_temp}
    T = \frac{\left|f'(v_+)\right| e^{-\chi(v_+)/2}}{4 \pi} \;.
\ee
Thus, noting that the solution at the horizon is AdS$_4$-RN (see Figure \ref{fig:ch4_3_estar_metricfuncts_all}), the temperature of the electron cloud solution reads
\be
    \label{eq:ch4_3_temp}
    T = \frac{6-\hat{q}^2}{8\pi\hat{v}_+} = \frac{6-\hat{q}^2}{8\pi} \;,
\ee
where we will work with the non-extremal $\hat{q}^2 < 6$ case. Additionally, looking at~\eqref{eq:ch4_3_scaling} and~\eqref{eq:ch4_3_nearbdy_A}, one can identify the physical parameter determining a unique background of the EC system to be $T/\hat{\mu}$ (invariant under dilatations), where $\hat{\mu}$ is the rescaled chemical potential of the boundary field theory given by $\hat{\mu} = \hat{h}(\hat{v})\rvert_{\hat{v}=0}$. We can then set the initial conditions at the horizon, namely $f'(\hat{v}_+)$, $\hat{h}'(\hat{v}_+)$, $\chi(\hat{v}_+)$ and $\chi'(\hat{v}_+)$, by their RN expressions above.

In Figure~\ref{fig:ch4_3_estar_metricfuncts_all}, the numerical solutions to the system are shown for a specific choice of the input parameters. We can see how the metric and gauge functions start deviating from the RN solution as one enters the electron cloud region. That is to say, the fermion fluid backreacts on the geometry of spacetime. Note that $f(\hat{v})\rightarrow 1$ when $\hat{v}\rightarrow 0$ as is expected by the requirement that spacetime be asymptotically AdS, but $\chi(\hat{v})$ does not tend to zero at the same limit. As it can be inferred from ansatz~\eqref{eq:EC_metric_ansatz}, this only affects the normalization of time. We could explicitly construct a correctly AdS normalized solution by exploiting a symmetry of the system given by
\be
    \label{eq:ch4_3_symmetry}
    \chi \, \rightarrow \, \chi - \chi_0 \;, \quad h \, \rightarrow \, e^{\chi_0/2}\,h \;,
\ee
where $\chi_0\equiv\chi(0)$, which can be checked directly from the equations of motion. It is also a good cross-check that $T/\hat{\mu}$ is invariant under~\eqref{eq:ch4_3_symmetry}.

\begin{figure}[t]
    \centering
    \includegraphics[scale=0.43]{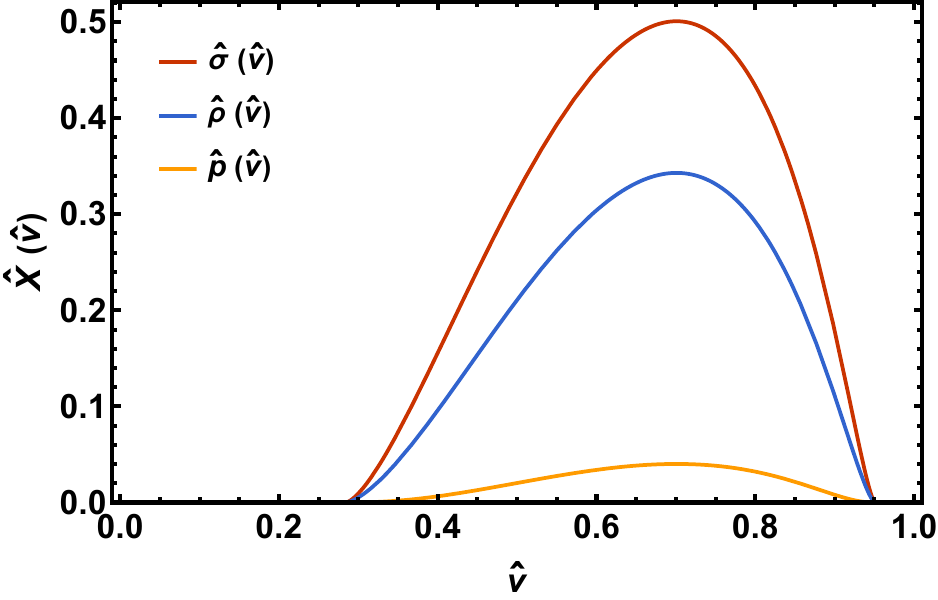}
    \caption{Numerical solutions to the properties of the fluid from the boundary ($\hat{v} \rightarrow 0$) to the outer event horizon ($\hat{v}_+ = 1$) in the EC system. This solution corresponds to $T/\hat{\mu} \approx 0.01$ and values for the input parameters: $\hat{\beta}=10$, $\hat{m}=0.55$ and $\hat{q} \simeq 2.23\,$.}
    \label{fig:ch4_3_estar_fluidfuncts_all}
\end{figure}

The fluid properties for the same choice of parameters are displayed in Figure~\ref{fig:ch4_3_estar_fluidfuncts_all}. These functions are identically zero outside the electron cloud region. From this plot, it is clear that there are two contributions to the total electric charge of the system: one coming from the charge of the black hole, $\hat{q}$, and another one resulting from integrating the non-zero charge density of the fermion fluid, $\hat{\sigma}(\hat{v})$. In addition, one can also match the geometry at the outer edge of the electron cloud toward the boundary to that of a planar AdS$_4$-RN black hole with charge equal to the total charge from both contributions, and a different normalization of time~\cite{Puletti:2010de}.

An important feature of this system we have not addressed is that it exhibits a third-order phase transition at a critical temperature of the black hole, $T_c$. For values of the temperature $T>T_c$, the electron cloud no longer has support in any region of the bulk and, thus, the solution shifts smoothly to the exact AdS$_4$-RN form. We are interested in studying the system subject to the backreaction from the fermion fluid, hence we work with $T<T_c$.\footnote{If the condition~\eqref{eq:ch4_3_fluidcond} is met for a non-zero region in the bulk, an electron cloud is formed, hence $T < T_c$.} For a detailed discussion of this phenomenon, we refer to~\cite{Hartnoll:2010ik}.

\begin{figure}[t]
    \centering
    \includegraphics[scale=0.8]{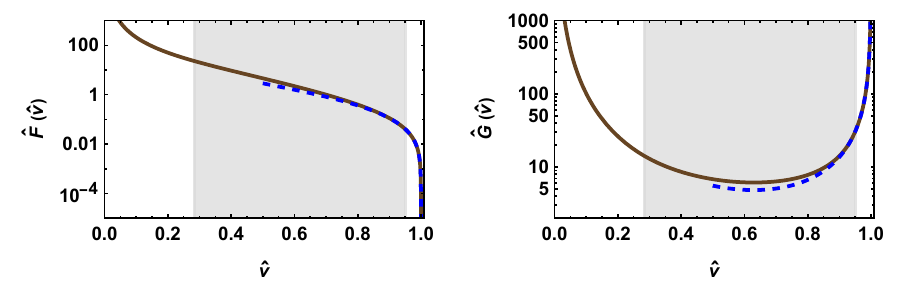}
    \caption{Numerical solutions to the metric functions, $\hat{F}(\hat{v})$ and $\hat{G}(\hat{v})$, defined as~\eqref{eq:ch4_3_ansatz_id1}, \eqref{eq:ch4_3_ansatz_id2}, from the boundary ($\hat{v} \rightarrow 0$) to the outer event horizon ($\hat{v}_+ = 1$) in the EC system. The region where the electron cloud is located is shown by the shaded gray area, and the blue-dashed line denotes the inner AdS$_4$-RN. This solution corresponds to $T/\hat{\mu} \approx 0.01$ and values for the input parameters: $\hat{\beta}=10$, $\hat{m}=0.55$ and $\hat{q} \simeq 2.23\,$.}
    \label{fig:ch4_3_estar_metricfuncts_check}
\end{figure}

Lastly, as a means to verify our results with Section 2 of~\cite{DiNunno:2021eyf} (where the same input parameters are employed), we identify the relationship between our ansatz and theirs 
\be
    \label{eq:ch4_3_ansatz_juan}
    ds^2 = -\hat{F}(\hat{v})d\hat{t}^2 + \hat{G}(\hat{v})d\hat{v}^2 + \frac{1}{\hat{v}^2}(d\hat{x}^2 + d\hat{y}^2) \;, \quad \hat{A}=\hat{H}(\hat{v})d\hat{t} \;.
\ee
That is,
\bea
    \hat{F}(\hat{v}) &=& \frac{1}{\hat{v}^2}f(\hat{v})e^{-\chi(\hat{v})} \;, \label{eq:ch4_3_ansatz_id1} \\
    \hat{G}(\hat{v}) &=& \frac{1}{\hat{v}^2f(\hat{v})} \;, \label{eq:ch4_3_ansatz_id2} \\
    \hat{H}(\hat{v}) &=& \hat{h}(\hat{v}) \;. \label{eq:ch4_3_ansatz_id3}
\eea
We plot the quantities corresponding to equations~\eqref{eq:ch4_3_ansatz_id1} and~\eqref{eq:ch4_3_ansatz_id2} in Figure~\ref{fig:ch4_3_estar_metricfuncts_check}, retrieving the correct profiles.

\section{Relevant deformations of the electron cloud \& Kasner interiors\label{sec3}}

Starting with~\cite{Frenkel:2020ysx}, a number of studies have explored the interior geometries of AdS black holes subject to relevant scalar deformations in their dual CFTs. Similarly, in this section, we proceed to solve the EC system under the same kind of deformation, examining the induced holographic RG flow in the full geometry. 
We find that the inner Cauchy horizon characteristic of a charged black hole is removed by the deformation, in agreement with existing horizon theorems for hairy black holes \cite{Hartnoll:2020rwq,Cai:2020wrp,An:2021plu,Arean:2024pzo}.
Instead, the interior geometry rapidly tends to a Kasner singularity, rendering a positive $p_t$ Kasner exponent related to the collapse of the Einstein-Rosen bridge.

\subsection{Equations of motion and numerical solutions}
We consider a $(3+1)$-dimensional Einstein-Maxwell-scalar system coupled to the electron cloud. The corresponding action reads
\be
    \label{eq:ch5_1_action}
    S = \int d^4x \sqrt{-g}(\pazocal{L}_{EH} + \pazocal{L}_{M} + \pazocal{L}_{F} + \pazocal{L}_{\phi}) \; ,
\ee
where the respective Lagrangian densities are \eqref{eq:lag_EH}, \eqref{eq:lag_Fsquared}, \eqref{eq:lag_fluid} and\footnote{The mass of the scalar field $m$ is not to be mistaken with the dimensionless mass of the fermions in the EC, $\hat{m}$.}
\be
    \label{eq:L_tot}
    \pazocal{L}_{\phi} = -\frac{1}{2}\partial^{\mu}\phi \partial_{\mu}\phi -\frac{1}{2} m^2\phi^2 \;.
\ee
The equations of motion consist of Einstein's equations, Maxwell's equations for a static fluid and the Klein-Gordon equation 
\be
    \label{eq:ch4_1_EOMs}
    G_{\mu\nu}-3g_{\mu\nu} = \kappa^2(T_{\mu\nu}^{M}+T_{\mu\nu}^{F}+T_{\mu\nu}^{\phi}) \;, \quad \nabla^{\nu}F_{\mu\nu} = e^2 J_{\mu}\;, \quad (\Box - m^2) \phi = 0 \;,
\ee
where the Maxwell's and fluid's stress-energy tensors are \eqref{eq:strener_M}, \eqref{eq:strener_F}, respectively, and $J_\mu$ is given by \eqref{eq:currentdens}. In addition, the scalar field's stress-energy tensor reads
\be
    \label{eq:Tmunu_phi}
    T_{\mu \nu}^{\phi} = \partial_{\mu}\phi\partial_{\nu}\phi - g_{\mu \nu}\(\frac{1}{2}\partial^{\mu}\phi \partial_{\mu}\phi +\frac{1}{2} m^2\phi^2\)\;.
\ee

Unlike the previous section, we do not work with the rescaled coordinates~\eqref{eq:ch4_3_scaling}, in which the horizon radius is fixed to $\hat{v}_+=1$, as we will need to vary $v_+$ to compute some observables in the following section. In spite of that, we follow the same considerations we did in the EC geometry regarding the treatment of the fluid, the use of dimensionless quantities and the equation of state. Therefore, to be consistent with these choices, we rescale $\phi = \frac{1}{\kappa}\hat{\phi}$. The scaling dimension of the relevant scalar operator in the CFT is chosen to be $\Delta = 2$, which corresponds to $m^2 = -2\,$ for the dual scalar field's mass satisfying $m^2 = \Delta(\Delta - 3)$.

The four independent equations of motion are in IEF coordinates \eqref{eq:ch4_3_ansatz_EF}
\bea
    h'' + \frac{1}{2}\chi'\,h' - \frac{\hat{\sigma}\,e^{-\chi/2}}{v^3\sqrt{f}} &=& 0 \;, \label{eq:ch5_1_eom_1}\\
    v^4 e^{\chi}\,h'^{\,2} + f \left ( 6+v^2(\hat{\phi}')^2 \right ) + 2\hat{\rho} - 2\left ( 3 + \hat{\phi}^{\,2} + vf' \right )  &=& 0 \;, \label{eq:ch5_1_eom_2}\\
    \frac{\chi'}{v} - (\hat{\phi}')^2 - \frac{h\,\hat{\sigma}\,e^{\chi/2}}{vf^{3/2}} &=& 0 \;, \label{eq:ch5_1_eom_3} \\
    \hat{\phi}\rq{}\rq{} + \left ( \frac{f'}{f} - \frac{2}{v} - \frac{\chi\rq{}}{2} \right ) \hat{\phi}\rq{} + \frac{2}{v^2 f} \hat{\phi} &=& 0 \;. \label{eq:ch5_1_eom_4}
\eea
To numerically solve these equations both in the exterior and interior of the black hole, we use a `two patch' method; we integrate from $v=v_+ - \epsilon$ backwards to the boundary $v=\epsilon$, and from $v=v_+ + \epsilon$ forward to the singularity $v=v_{\text{max}}$ (for some comparatively big $v_{\text{max}}>v_+ + \epsilon$). Integrating in these two patches, we avoid any numerical issues coming from $f(v)$ vanishing at $v_+$ in some denominators of the equations of motion. Consequently, we need to fix the initial conditions at the horizon $v_+$. We know that $f(v_+)=0$ and $h(v_+)=0$. Additionally, expanding the equations above around $v = v_+$ up to leading order, we arrive at the following set of constraints
\bea
    \hat{\phi}(v_+) &=& \sqrt{-3-f'(v_+)v_+ + \frac{1}{2}e^{\chi(v_+)}h'(v_+)^2v_{+}^4} \;, \label{eq:ch5_1_init_1} \\
    \hat{\phi}'(v_+) &=& -\frac{\sqrt{2}\sqrt{-6-2f'(v_+)v_+ + e^{\chi(v_+)}h'(v_+)^2 v_{+}^4}}{f'(v_+)v_{+}^2} \;, \label{eq:ch5_1_init_2} \\
    \chi'(v_+) &=& \frac{2\left ( -6 -2f'(v_+)v_+ + e^{\chi(v_+)}h'(v_+)^2 v_{+}^4\right )}{f'(v_+)^2v_{+}^3} \;. \label{eq:ch5_1_init_3}
\eea

Regarding $f'$ and $h'$, we set them to be those of the AdS$_4$-RN solution plus an arbitrary perturbation to account for the scalar deformation, i.e., $f'(v_+) = f_{\text{\tiny{RN}}}'(v_+) - f_{1,\,pert}\,$ and $\,h'(v_+) = h_{\text{\tiny{RN}}}'(v_+) - h_{1,\,pert}$. This is justified in the same way that the value of the horizon radius $v_+$ is arbitrary: these quantities are not invariant under dilatations. The last initial condition needed is the value of $\chi(v)$ at the horizon. Thanks to the symmetry~\eqref{eq:ch4_3_symmetry}, we can set this to be $\chi(v_+)=0$ and then choose the correct normalization of the time coordinate by shifting the function and rescaling the gauge potential accordingly.

It is crucial to note that there cannot be another electron cloud inside the black hole. That is so because in the interior $f(v)<0$, hence the bulk chemical potential defined in~\eqref{eq:ch4_3_muloc_def} becomes imaginary. This implies that the backreaction from the cloud geometry comes solely from the exterior. Without deformation, we know the interior consists of an AdS$_4$-RN geometry, presenting an inner Cauchy horizon followed by a timelike singularity. We want to study the effects the relevant deformation has on this regime (named in~\cite{Caceres:2022smh} as the `trans-IR' regime).

In this case there are two independent dimensionless ratios determining a unique solution to the background, $T/\hat{\phi}_0$ and $T/\mu$, where $\hat{\phi}_0=\lim_{v\to0}v^{-1}\hat{\phi}(v)$ is the source for the CFT scalar operator. For the system to be at low temperatures, hence exhibit an electron cloud, both of these ratios have to be $T/\hat{\phi}_0 \ll 1\,,\,T/\mu \ll 1$. We constrain the solutions to this regime.

\begin{figure}[t]
    \centering
    \includegraphics[scale=0.45]{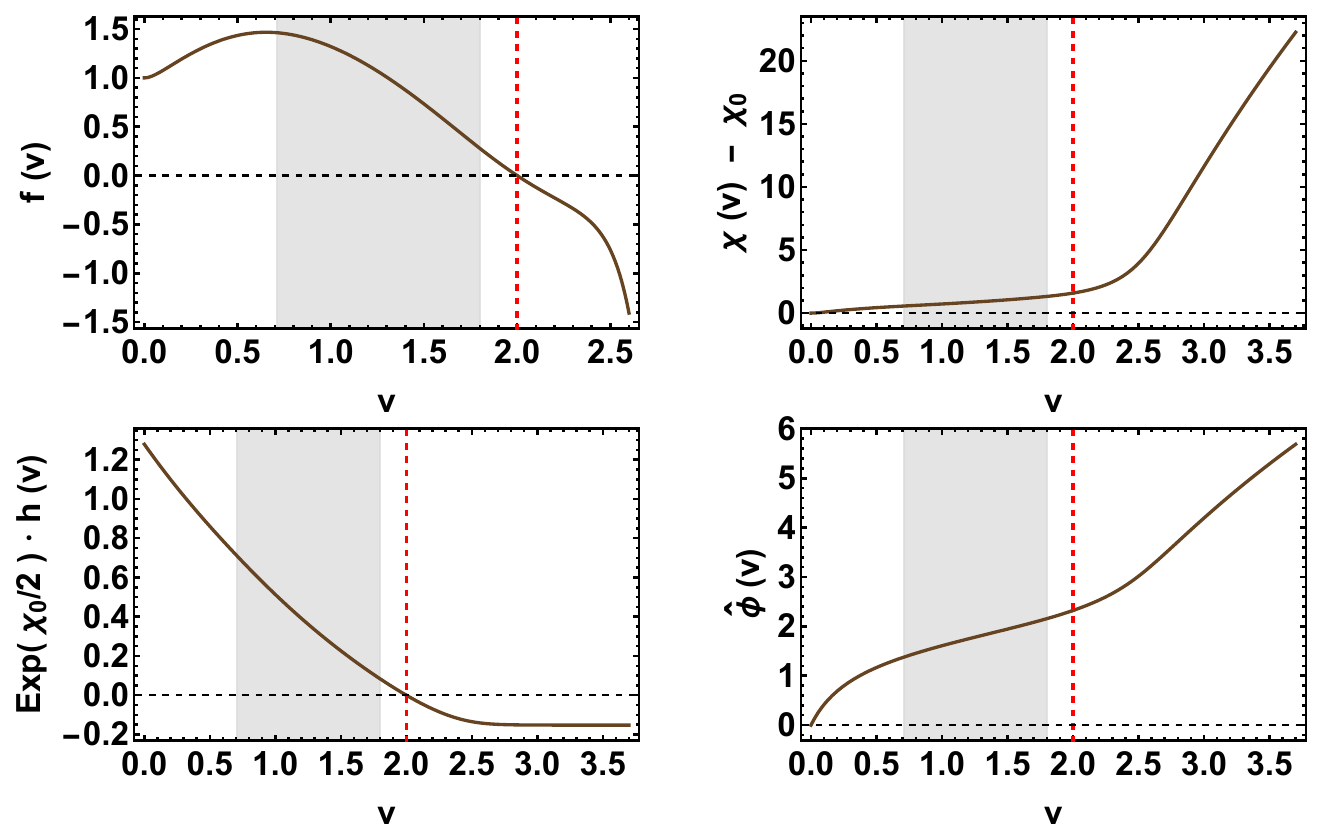}
    \caption{Numerical solutions to the metric functions $f(v)$ (upper left) and $\chi(v)$ (upper right), gauge field $h(v)$ (lower left) and scalar field $\hat{\phi}(v)$ (lower right), with the correct normalization of time, from the boundary into the interior in the deformed EC system. The region where the electron cloud is located is shown by the shaded gray area, and the red-dashed line indicates the horizon radius, $v_+=2$. This solution corresponds to $T/\hat{\phi} \approx 0.009$, $T/\mu \approx 0.036$ and values for the input parameters: $\hat{\beta}=10$, $\hat{m}=0.55$ and $\hat{q} \simeq 2.23\,$.}
    \label{fig:ch5_1_estarscalar_metricfuncts_all}
\end{figure}

Figure~\ref{fig:ch5_1_estarscalar_metricfuncts_all} shows the results for a specific background of the deformed EC system. In these plots, we make use of the mentioned symmetry to display the correctly normalized metric functions, $\Tilde{\chi}(v)\equiv \chi(v) - \chi_0\,$ and $\,\Tilde{h}(v) \equiv e^{\chi_0/2}h(v)$. We find that the integration is smooth through the horizon, and $f$ remains negative asymptotically after $v=v_+$. This indicates that the inner Cauchy horizon is removed by the relevant scalar deformation. Closely following the proof constructed in \cite{Hartnoll:2020rwq}, we can see this analytically starting by rewriting the scalar field's equation of motion \eqref{eq:ch5_1_eom_4} multiplied by $f(v)$ as
\be
\label{eq:noinnerhor_proof_eom}
v^2 e^{\chi/2} \frac{d}{dv}\(e^{-\chi/2} v^{-2} f \hat{\phi}'\)+\frac{2}{v^2}\hat{\phi}=0 \,.
\ee
Then, assuming the existence of an inner horizon $v=v_-$ where $f(v_-)=0$, one has
\be
\label{eq:noinnerhor_proof}
0=\int_{v_+}^{v_-}\frac{d}{dv}\(e^{-\chi/2} v^{-2} f \hat{\phi}'\hat{\phi}\) dv = \int_{v_+}^{v_-} \frac{e^{-\chi/2}}{v^4}\left[ -2 \hat{\phi}^2 + v^2f(\hat{\phi}')^2\right ] dv \,,
\ee
where the first equality follows from $f(v_+)=f(v_-)=0$, and we use \eqref{eq:noinnerhor_proof_eom} in the second equality. Thus, noting that $f<0$ for $v_+<v<v_-$, the integrand in the rightmost side of \eqref{eq:noinnerhor_proof} is negative in the range of integration; we arrived at a contradiction unless $\hat{\phi}=0$. In words, when the massive scalar field with $m^2 \leq 0$ (here $m^2=-2$) is turned on, there cannot be an inner Cauchy horizon.

\subsection{Near horizon criticality}
Previous works on electron stars, and electron cloud geometries~\cite{Hartnoll:2009ns,Hartnoll:2010gu} showed that they develop a non-relativistic Lifshitz scaling symmetry at low temperatures in the IR. Said symmetry is characterized by a dynamical critical exponent, $z$, such that the theory is invariant under~\cite{Kachru:2008yh,Azeyanagi:2009pr,Taylor:2015glc,Tarrio:2011de}
\be
    \label{eq:ch5_2_Lifshitz_scaling}
    t \, \rightarrow \, \lambda^z\,t \;, \quad x^{a} \, \rightarrow \, \lambda\,x^a \;,
\ee
where $z\neq 1$. It places time and spatial coordinates on different footings, hence breaking Lorentz invariance. This scaling is found in many fixed points determining phase transitions of condensed matter systems. Particularly, the holographic Lifshitz geometry serves as a gravitational dual to the physics of strange metals (non-Fermi liquids) at quantum critical points.\footnote{One consequence of the Lifshitz scaling is that, at low temperature, the specific heat exhibits scaling behavior as
$C_v\sim T^{d/z}$, where $d$ is the number of boundary space dimensions. This implies that asymptotically Lifshitz black holes provide a promising holographic description of non-relativistic Fermi liquids when $z = d$. A variant of these models, which introduces an additional `hyperscaling violating' exponent $\theta$, allows for the realization of Fermi liquid behavior for arbitrary values of $z(d)$~\cite{Dong:2012se,Narayan:2012hk,Edalati:2012tc,Gath:2012pg,Edalati:2013tma,Pedraza:2018eey}.}

The IR of our deformed EC system corresponds to the black hole's horizon. This hints us that properties depending on the horizon must be somehow related to the $z$ exponent. One such property is the entropy $S$ of the black hole, given by the Bekenstein-Hawking law
\be
    \label{eq:ch1_entropyarea}
    S = \frac{A}{4 G_N} \;,
\ee
where $G_N$ is Newton's gravitational constant in four spacetime dimensions and natural units are employed. From the ansatz~\eqref{eq:EC_metric_ansatz}, the area of the horizon is simply $A = \frac{1}{v_{+}^2}\int_{\mathbb{R}^2}dx\,dy$, which diverges due to its planar topology. To circumvent this, we work with the entropy density 
\be
    \label{eq:ch5_2_entropydens}
    s = \frac{1}{4v_{+}^2} \;,
\ee
where we have taken $G_N = 1$.
\begin{figure}[t]
    \centering
    \includegraphics[scale=0.55]{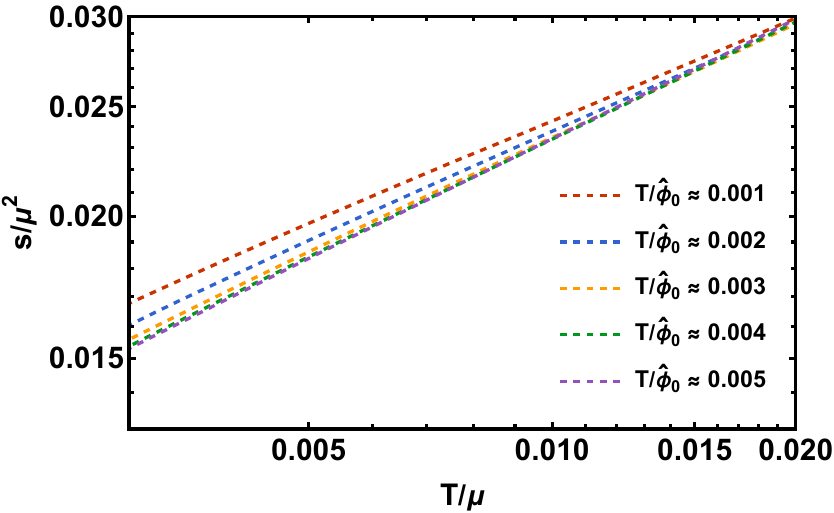}
    \caption{Numerical solution to the entropy density as a function of temperature for different fixed values of $T/\hat{\phi}_0$. The low-temperature fitting of these results leads to (from top to bottom): $z \approx 6.30,\,5.74,\,5.60,\,5.56,\,5.57\;$. This solution corresponds to values for the input parameters: $\hat{\beta}=10$, $\hat{m}=0.55$ and $\hat{q} \simeq 2.23\,$.}
    \label{fig:ch5_2_s_T}
\end{figure}

We want to find another quantity that depends on $v_+$ and $z$ to relate to $s$. The near-horizon geometry must have a correction in the metric's time component, i.e.
\be
    \label{eq:ch5_2_Lifshitz_BH}
    ds^2 = - \frac{1}{v^{2z}}f(v)\,dt^2 + \frac{1}{v^2 f(v)}dv^2 + \frac{1}{v^2}d\Vec{x}^{\,2} \;.
\ee
Consider, for instance, a $(3+1)$-dimensional AdS-Schwarzschild black hole such that $f(v) = 1 - \left ( \frac{v}{v_+} \right )^3$. Then, the temperature derived from the Lifshitz-invariant metric above via~\eqref{eq:kasner_BH_temp} is
\be
    \label{eq:ch5_2_temp_LSchw}
    T = \frac{\left | f'(v_+) \right |}{4\pi} v_{+}^{1-z} = \frac{3}{4\pi}\frac{1}{v_{+}^z} \;.
\ee
In short, we have $s \propto v_{+}^{-2}$ and $T \propto v_{+}^{-z}$ such that $s \propto T^{2/z}$. This applies to any planar AdS-black hole (with $d=3$) presenting a Lifshitz symmetry at the horizon, since we are only using the dimensionality of the entropy density and temperature as arguments. Thus, we have found a way to extract the dynamical critical exponent $z$ of the IR geometry. We show the results of this dependence in Figure~\ref{fig:ch5_2_s_T} for several fixed values of $T/\hat{\phi}_0$. To construct this plot, we solve many backgrounds varying the parameter $\hat{q}$, which changes $T/\mu$, while adjusting the horizon radius $v_+$ to keep $T/\hat{\phi}_0$ fixed.

\subsection{Kasner interiors}
We would like to understand how the black hole's interior is affected by the deformation, given that, without it, its geometry is that of AdS$_4$-RN. We find that the inner horizon is destroyed and the scalar field grows logarithmically as it approaches the singularity. The divergent behaviour of the fields near the singularity is given by
\be
    \hat{\phi}(v) \sim c \log v\;,\;\;\chi(v) \sim 2c^2\log v+\chi_0 \;,\;\;f(v) \sim -f_1 v^{3+c^2}
\ee
where $c$ is a constant.

\begin{figure}[t]
    \centering
    \includegraphics[scale=0.5]{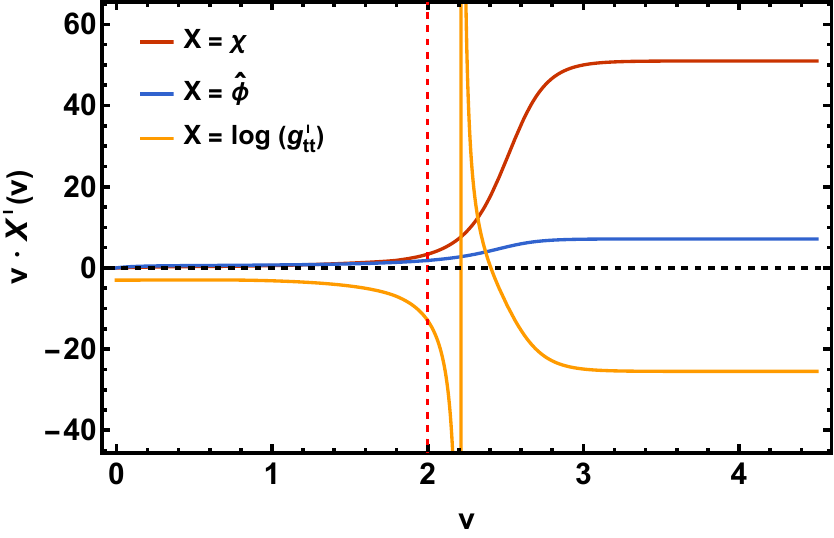}
    \includegraphics[scale=0.5]{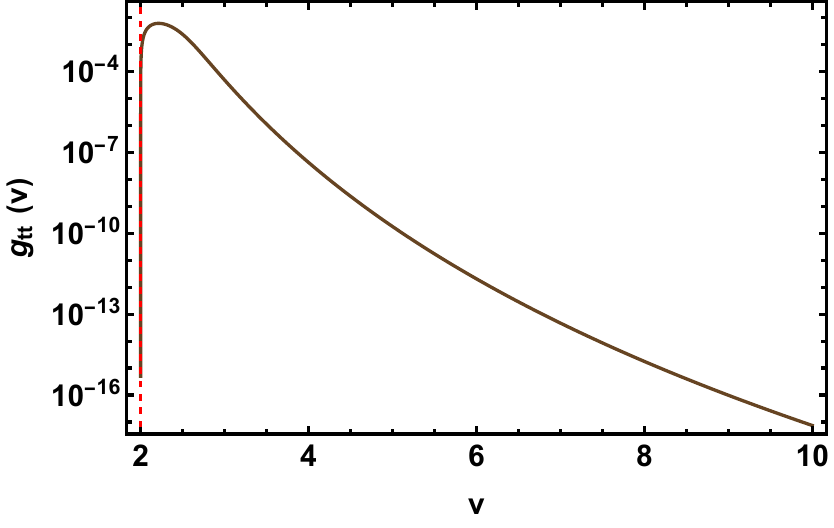}
    \caption{\textbf{Left:} Near-singularity behavior of spacetime in the DEC system. The plotted quantities tend to constant values as $v\rightarrow \infty$, demonstrating that the geometry adopts a more general Kasner form. \textbf{Right:} Time component of the metric~\eqref{eq:EC_metric_ansatz} as a function of the holographic radial coordinate $v$ inside the black hole in the DEC system. The red-dashed line indicates the horizon radius at $v_+=2$. These solutions correspond to $T/\hat{\phi} \approx 0.009$, $T/\mu \approx 0.036$ and values for the input parameters: $\hat{\beta}=10$, $\hat{m}=0.55$ and $\hat{q} \simeq 2.23\,$.}
    \label{fig:ch5_1_estarscalar_gtt}
\end{figure}
We have all the necessary tools to explicitly check that, near the singularity, the metric adopts a Kasner form under a change of coordinates given by $v = \tau^{-\frac{2}{3+c^2}}$
\be
\label{eq:schw_to_kasner_metric}
ds^2 \sim -d\tau^2 + \tau^{2p_t}dt^2 + \tau^{2p_x}(dx^2 + dy^2) \; \;, \quad \hat{\phi}(v) \sim -\,\sqrt{2}p_{\hat{\phi}} \log \tau \;.
\ee
The Kasner exponents are 
\be
    \label{eq:kasner_exponents_t_x}
    p_t = \frac{c^2-1}{3+c^2} \;, \quad p_x = \frac{2}{3+c^2} \;\;,\;\; p_{\hat{\phi}}= \frac{2\sqrt{2}c}{3+c^2}\;,
\ee
satisfying $p_t + 2p_x = 1$ and $(p_t)^2+2(p_x)^2+(p_{\hat{\phi}})^2=1$. This corresponds to a generalized Kasner cosmology, i.e. a Kasner universe in the presence of a matter field $\hat{\phi}$.

\begin{figure}[t]
    \centering
    \includegraphics[scale=0.55]{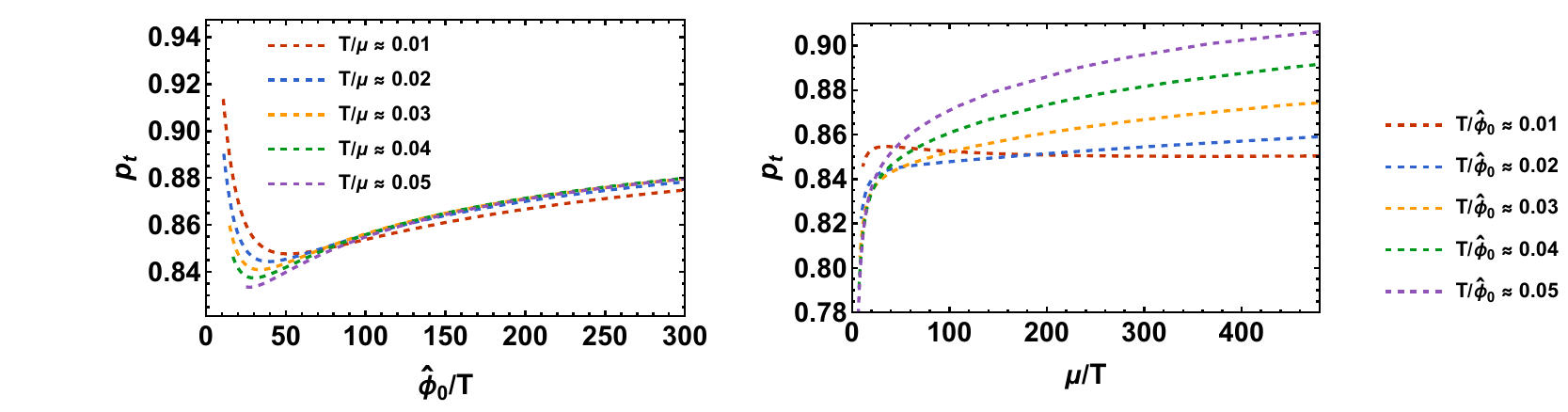}
    \caption{Numerical solution to the Kasner exponent $p_t$ as a function of $\hat{\phi}_0/T$ for different fixed values of $T/\mu$ (left) and as a function of $\mu/T$ for different fixed values of $T/\hat{\phi}_0$ (right), in the deformed EC system. These solutions correspond to values for the input parameters: $\hat{\beta}=10$, $\hat{m}=0.55$ and $\hat{q} \simeq 2.23\,$.}
    \label{fig:ch5_1_kasner_pt}
\end{figure}

Given that there are three Kasner exponents and two constraints, we explore the dependence of $p_t$ with the parameters that determine a unique background $T/\hat{\phi}_0$ and $T/\mu$ in Figure~\ref{fig:ch5_1_kasner_pt}. The most striking feature is the positiveness of $p_t$, $0\leq p_t<1$, contrary to the studied deformed AdS-Schwarzschild case~\cite{Frenkel:2020ysx}. Moreover, $p_t$ is close to $1$ such that $p_x = \frac{1}{2}(1-p_t)$ is close to $0$. We identify this behavior as the collapse of the Einstein-Rosen bridge, since the transversal part of the metric shrinks in the new interior Kasner spacetime. For a more detailed explanation of this phenomenon, we refer to~\cite{Hartnoll:2020rwq}.

\begin{figure}[t]
    \centering
    \includegraphics[width=0.4\columnwidth]{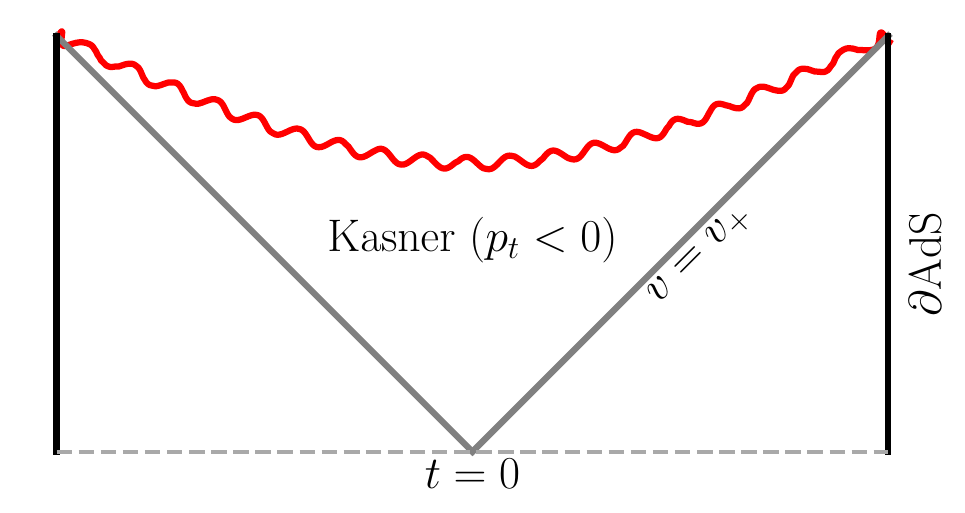}\hspace{1cm}
    \includegraphics[width=0.4\columnwidth]{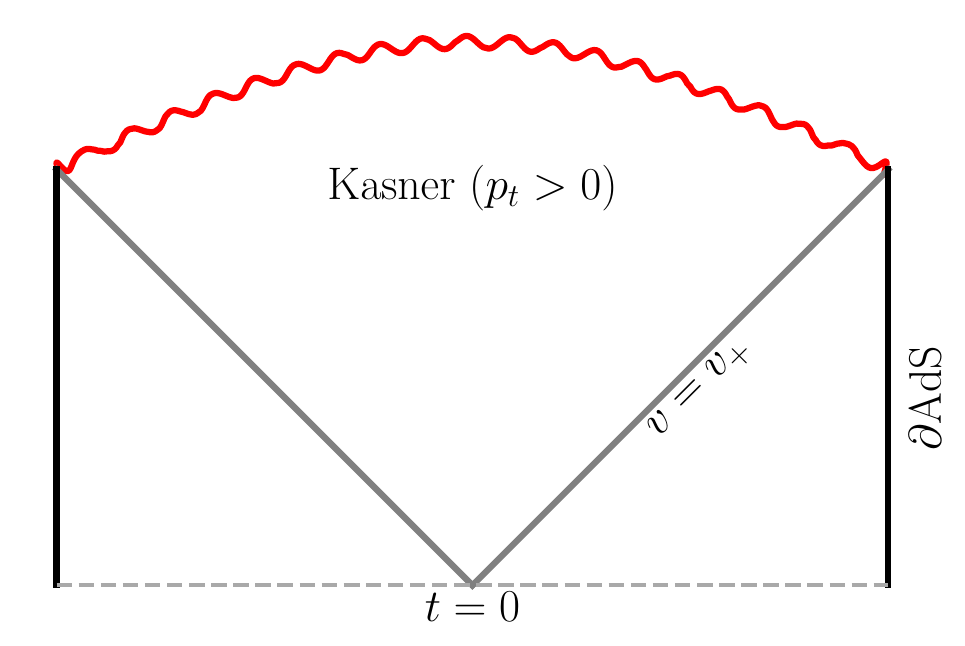}
    \caption{Penrose diagrams for black hole solutions with Kasner singularities are shown for $p_t<0$ (left) and $p_t>0$ (right). In the $p_t<0$ case, the singularity bulges downward, resembling the behavior in Schwarzschild-AdS black holes. Conversely, in the $p_t>0$ case, the singularity bulges upward, approaching the would-be Cauchy horizon.}
    \label{fig:penrose}
\end{figure}

It is important to note that the positive-$p_t$ behavior complicates direct probing of the singularity \cite{Hartnoll:2020rwq}. For instance, spacelike geodesics become trapped near the horizon and fail to penetrate into the deep interior, contrasting with the behavior observed for negative-$p_t$ solutions \cite{Frenkel:2020ysx}. This feature can be intuitively understood through the corresponding Penrose diagrams shown in Figure~\ref{fig:penrose}. When $p_t$ is negative, the singularity bulges downward, as typically seen in Schwarzschild-AdS black holes. In contrast, for positive $p_t$, the singularity bulges upward, approaching the would-be Cauchy horizon. Consequently, it becomes increasingly difficult for spacelike probes to access the region near the singularity. In the next section, however, we will demonstrate that certain CFT probes can still access the near-singularity geometry. These include the thermal $a$-function and a variant of the geodesics that can be associated with charged correlators. Additionally, we will examine other observables that, while not directly probing the singularity, provide valuable insights into the black hole interiors and the corresponding RG flows.

\section{Probing the RG flows\label{sec4}}
Now that we have solved the deformed electron cloud system, we can extract some interesting observables and probes of its full geometry.

In this section, we first examine the thermal generalization of the so-called holographic $a$-function and explain its relevance to the RG flow. Then, we compute the two-point correlation functions, complexity, and entanglement entropy velocities via their holographic prescriptions. Lastly, we introduce the butterfly velocity and explore its bounds in this system.

\subsection{Thermal $a$-function and number of degrees of freedom}
The original $c$-theorem~\cite{Zamolodchikov:1986gt} states that, for any $2$-dimensional QFT, there exists a function on the space of couplings $\lambda^i$, $c(\lambda^i)$, such that it monotonically decreases under a renormalization group transformation given by the beta function $\beta(\lambda(\mu)) \equiv \mu \frac{\partial \lambda}{\partial \mu}$. Moreover, this function only has constant values at the RG fixed points, where it is equivalent to the central charge of the corresponding CFT. The latter, in turn, can be interpreted as a measure of the number of degrees of freedom of the theory. Therefore, from the Wilsonian perspective, the monotonic decrease comes from integrating out the UV degrees of freedom.

The holographic $a$-function~\cite{Myers:2010tj,Freedman:1999gp} represents an extension of this theorem to an arbitrary number of dimensions. It decreases monotonically along the holographic radial coordinate $v$, which describes the flow from the UV (conformal boundary) to the IR. This behavior reflects the RG flow of the field theory, consistent with the standard UV/IR connection in AdS/CFT~\cite{Susskind:1998dq,Peet:1998wn,Hatta:2010dz,Agon:2014rda}. The thermal $a$-function, $a_T$, we explore here is just a further generalization accounting for a thermal state in the CFT, i.e. a black hole spacetime in the bulk.

We compute the $a_T$ function of our system following \cite{Caceres:2022smh}
\be
    \label{eq:ch6_1_aT}
    a_T(v) = \frac{\pi^{3/2}}{\Gamma(\frac{3}{2})\,l_{p}^2}\,e^{-\Tilde{\chi}(v)} \;,
\ee
where we have set $d=3$ and we mind to use the correct time normalized metric function $\Tilde{\chi}(v)$.

Another interesting quantity to compute is the derivative of $a_T$ with respect to the exterior and (analytically continued) interior radial coordinates given in the same reference, which they globally denote $E$. As we mentioned earlier, this function is only stationary at the RG fixed points, hence providing a direct method of finding them. Applying the chain rule, one arrives at \cite{Caceres:2022smh}
\be
\label{eq:ch6_1_dadE}
\frac{da_T}{dE} = 
\left\{
    \begin{array}{ll}
        \Tilde{\chi}'(v)\,a_T(v)\,v\sqrt{f(v)}\;, & \quad v \leq v_+ \\
        -\Tilde{\chi}'(v)\,a_T(v)\,v\sqrt{\left | f(v) \right |}\;, & \quad v > v_+
    \end{array}
\right. \;.
\ee
\begin{figure}[t]
    \centering
    \includegraphics[scale=0.55]{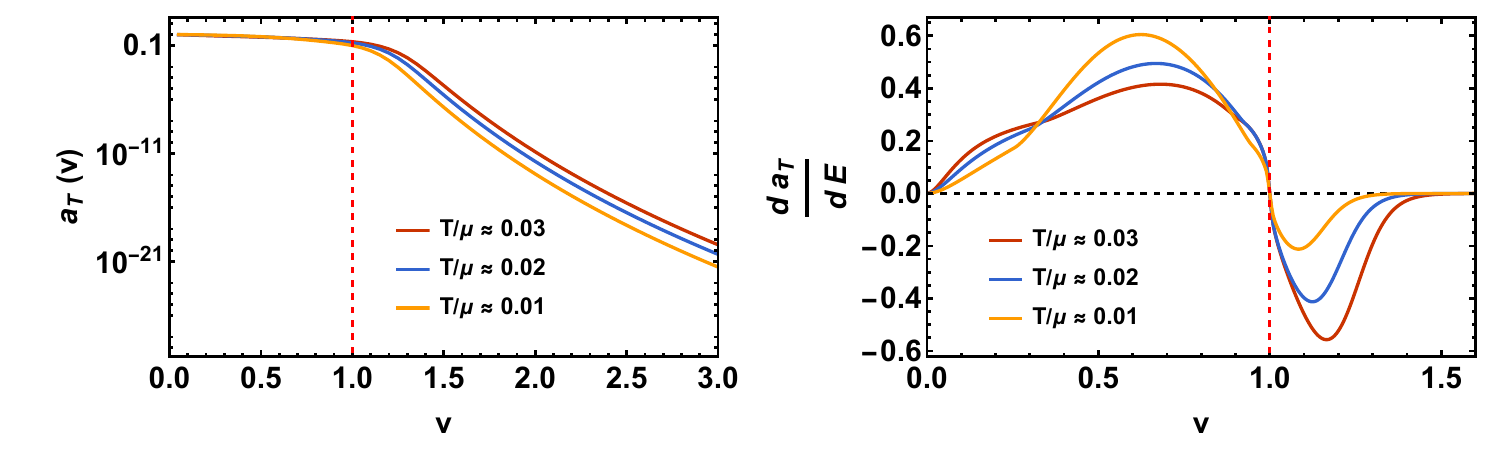}
    \caption{Monotonic $a_T$-function and its generalized derivative $da_T/dE$ in units of $2\pi l_{p}^{-2}$ for three different backgrounds with fixed $T/\hat{\phi}_0 \approx 0.01$ in the deformed EC system. The red dashed line indicates the black hole's horizon at $v_+ = 1\,$. The cloud input parameters are: $\hat{\beta}=10$, $\hat{m}=0.55$ and $\hat{q} \simeq 2.23\,$.}
    \label{fig:ch6_1_afunction}
\end{figure}
In Figure~\ref{fig:ch6_1_afunction}, we check that the $a_T$-function does indeed decrease monotonically along the holographic RG flow and tends to zero near the singularity, indicating a total loss of degrees of freedom. On the other hand, from the plot on the right, it is clear that $a_T$ is stationary at the boundary, the horizon and the singularity. Therefore, these represent three fixed points of the RG flow where the degrees of freedom `freeze out': the UV, IR and trans-IR fixed points respectively. The main difference with the results of~\cite{Caceres:2022smh} for the deformed AdS-Schwarzschild flows lies in the appearance of the trans-IR fixed point. We now see that we can attribute this distinction to the positive values of the Kasner $p_t$ exponent. 

To prove that, let us rewrite the near-singularity behaviors of the metric functions
\be
    \label{eq:ch6_1_nearsing}
    \Tilde{\chi}(v) \sim 2c^2 \log v \;, \quad f(v) \sim -f_1 v^{3+c^2} \;,
\ee
which, in turn, imply
\be
    \label{eq:ch6_1_nearsing_at}
    a_T(v) \sim v^{-2c^2} \;.
\ee
Now, substituting~\eqref{eq:ch6_1_nearsing} and~\eqref{eq:ch6_1_nearsing_at} into the $v>v_+$ part of~\eqref{eq:ch6_1_dadE} yields
\be
    \label{eq:ch6_1_nearsing_dadE}
    \frac{da_T}{dE} \sim -v^{\sigma} \;,
\ee
where $\sigma \equiv \frac{3}{2}(1-c^2)$. The sign of $\sigma$ is what determines whether there is a trans-IR fixed point or not. It is easier to write this exponent in terms of $p_t = \frac{c^2-1}{3+c^2}$ as 
\be
    \label{eq:ch6_1_sigma_pt}
    \sigma = \frac{6p_t}{p_t-1} \;.
\ee
As $p_t$ takes values between $0$ and $1$, $\sigma$ is negative and the $a_T$-function is stationary at $v\rightarrow \infty$. Also, the closer $p_t$ is to $1$, the quicker $da_T/dE$ goes to zero, in relation with the speed of the collapse of the Einstein-Rosen bridge.

\subsection{Correlators of heavy operators and geodesics}
Studies have demonstrated that spacelike geodesics are invaluable tools for probing the regions near a black hole's singularity. Through AdS/CFT correspondence, these geodesics are directly related to the two-point correlation functions of heavy boundary operators placed one on each asymptotically AdS boundary. For a given boundary operator $\mathcal{O}$ with conformal dimension $\Delta_\mathcal{O}$, the correlation function between two points $X$ and $Y$ on the boundary can be expressed in `first quantized' language as \cite{Balasubramanian:1999zv,Louko:2000tp}
\begin{equation}
   \braket{{\mathcal{O}}(X){\mathcal{O}}(Y)}=\int_{X}^{Y} d \mathcal{P} e^{-\Delta_{\mathcal{O}} L(\mathcal{P})}\,,
\end{equation}
where $\mathcal{P}$ represents all possible paths connecting $X$ and $Y$, and $L(\mathcal{P})$ is their lengths. For large conformal dimensions ($\Delta_{\mathcal{O}} \rightarrow \infty$), the path integral is dominated by the shortest geodesic, allowing us to approximate the correlation function as:
\begin{equation}
   \braket{{\mathcal{O}}(X){\mathcal{O}}(Y)}\approx\sum_{\text{geodesics}}e^{-\Delta_{\mathcal{O}} L(X,Y)}\approx e^{-\Delta_{\mathcal{O}} L_M(X,Y)} \,,
   \label{twopointc}
\end{equation}
where $L(X,Y)$ is regularized geodesic length between $X$ and $Y$, and $L_M(X,Y)$ is the minimal regularized geodesic length between $X$ and $Y$. 

We are concerned with calculating the 2-point correlation function between two symmetrically placed boundary points, one on each AdS boundary. Using the relation in (\ref{twopointc}), we analyze the spacelike geodesics anchored at those endpoints in the bulk. Because of the spatial symmetry in such configurations, these geodesics reside on a constant plane described by the coordinates $(t,v)$. The induced metric on such a plane is given by:
\begin{equation}
ds^2=\frac{1}{v^2}\left[-f(v)e^{-\chi(v)} dt^2 + \frac{dv^2}{f(v)}\right]\,.
\end{equation}
Any spacelike geodesics on this plane can be parameterized by $v = v(t)$, and their length functional can be expressed as:
\begin{equation}
    L=\int \frac{dt}{v}\left[-f(v)e^{-\chi(v)}+\frac{\dot{v}^2}{f(v)}\right]^{{1}/{2}}=\int dt \mathcal{L}\label{geol}\,,
\end{equation}
where $\mathcal{L}$ is the Lagrangian,
\begin{equation}
    \mathcal{L}=\frac{1}{v}\sqrt{-f(v)e^{-\chi(v)}+\frac{\dot{v}^2}{f(v)}}\,.
\end{equation}
Given that the Lagrangian does not explicitly depend on time, the energy $\mathcal{E}$ associated with the geodesic is conserved and can be defined as:
\begin{equation}
    \mathcal{E}=\dot{v}\frac{\partial\mathcal{L}}{\partial\dot{v}}-\mathcal{L}=\frac{f(v)e^{-\chi(v)}}{v\sqrt{-f(v)e^{-\chi(v)}+\frac{\dot{v}^2}{f(v)}}}\,.
    \label{constom}
\end{equation}
Substituting this expression back into the integral for the geodesic length in (\ref{geol}), we find the minimal geodesic length anchored at some boundary time slice $t=t_b(\mathcal{E})$,
\begin{equation}
    L=\frac{2}{|\mathcal{E}|}\int_{v_\epsilon}^{v_t} \frac{e^{-\chi/2}dv}{v^2\sqrt{1+\frac{f(v)e^{-\chi(v)}}{v^2\mathcal{E}^2}}}+2 \log v_\epsilon\,.\label{emptylength}
\end{equation}
We have introduced a counterterm to subtract the UV divergences arising from the AdS boundary. Here, $v_\epsilon$ represents the UV cutoff, and $v_t$ is the turning point of the geodesic, which is determined by the relation,
\begin{equation}
\mathcal{E}^2=g_{tt}(v_t)=-\frac{f(v_t)e^{-\chi(v_t)}}{v_t^2}\,.
\label{energyturn}
\end{equation}
The turning point $v_t$ can be used further to express the boundary time $t_b$ as a function $\mathcal{E}$,
\begin{equation}
    t_b=-\int_{0}^{v_t}\frac{dv}{\dot{v}}=-\int_{0}^{v_t}\frac{e^{\chi/2}dv}{f(v)\sqrt{1+{f(v)e^{-\chi(v)}}/(v \mathcal{E})^2}}\,.
    \label{boundarytime}
\end{equation}
With increasing energy $\mathcal{E}$ in (\ref{energyturn}), geodesics can probe deeper regions in the black hole interior. If this energy grows without bound, then the turning point $v_t$ can go arbitrarily close to the singularity. However, if there exists a maximum value for $\mathcal{E}$ at some critical radius $v_m$, then the geodesics are stuck at that radius and can not probe any deeper regions inside of the black hole. In our system, this is the case. We find that in DEC, we are not allowed to take $\mathcal{E}$ arbitrarily large and thus are unable to see any near singularity effects on the two-point correlation function.

To increase this maximum value for $\mathcal{E}$, we next consider the evaluation of two-point correlation functions of massive \emph{charged} operators, as the coupling of the charged particle with the Maxwell field could potentially enable it to probe the black hole's deep interior.\footnote{We are grateful to Miguel Montero for suggesting us to investigate correlators of charged operators.} In the saddle point approximation, it has been established that for heavy-charged operators inserted at $X$ and $Y$, the two-point correlation function is given by \cite{Giordano:2014kya},
\begin{equation}
\braket{\mathcal{O}(X)\mathcal{O}(Y)}\approx e^{-\Delta_{\mathcal{O}} S_{\text{on-shell}}(X,Y)} \,,
\end{equation}
where $S_{\text{on-shell}}$ is the on-shell action for a space like charged geodesic,
\begin{equation}
    S_{\text{on-shell}}=\int dt \bigg(\sqrt{g_{\mu\nu}\dot{x}^\mu\dot{x}^\nu}+q_E A_\mu\dot{x^{\mu}}\bigg)\bigg|_{\text{{on-shell}}}\label{actioncharged} \,,
\end{equation}
$q_E=q/m$ is the charge per unit mass, and $A_\mu$ is defined in (\ref{eq:EC_metric_ansatz}).\footnote{We recall that particles with charge $q/m>1$ are required to preserve weak cosmic censorship \cite{Crisford:2017gsb,Horowitz:2019eum}.}
In our setup, the action $S$ for symmetrically placed boundary-charged operators, one on each AdS boundary, is,
\begin{equation}
    S=\int dt \bigg\{ \frac{1}{v}\left[-f(v)e^{-\chi(v)}+\frac{\dot{v}^2}{f(v)}\right]^{{1}/{2}}+q_E \bigg(A_t(v)+ A_v(v) \dot{v}\bigg)\bigg \}=\int dt \mathcal{L} \,.
\end{equation}
Proceeding just like before, we define a new conserved quantity,
\begin{equation}
    \mathcal{E}=\dot{v}\frac{\partial \mathcal{L}}{\partial\dot{v}}-\mathcal{L}=\frac{f(v)e^{-\chi(v)}}{v\sqrt{-f(v)e^{-\chi(v)}+\frac{\dot{v}^2}{f(v)}}}-q_E A_t(v)
    \label{newenergy}
\end{equation}
\begin{figure}
    \centering
    \includegraphics[width=1\linewidth]{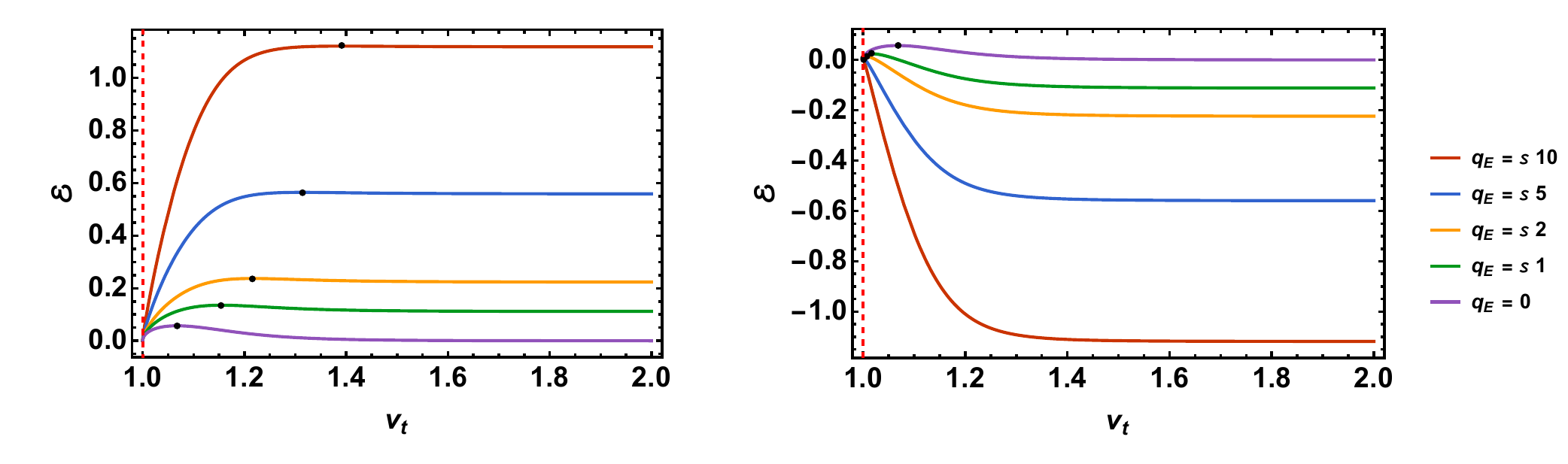}
    \caption{Energy of the spacelike geodesics $\mathcal{E}$ as a function of the turning point $v_t$. The maxima of these energies (black dots) represent the deepest points in the black hole's interior. The sign of the probe charge $s$ is positive ($s=+1$) in the left plot and negative ($s=-1$) in the right plot. The red-dashed line indicates the horizon radius at $v_+=1$. This solution corresponds to $T/\hat{\phi} \approx 0.009$, $T/\mu \approx 0.010$ and values for the input parameters: $\hat{\beta}=10$, $\hat{m}=0.55$ and $\hat{q} = 5.30\,$. }
    \label{energyvsturn}
\end{figure}
Substituting this expression of $\mathcal{E}$ in the on-shell action in (\ref{actioncharged}), we find,
\begin{equation}
    S_{\text{on-shell}}=2\int_{v_c}^{v_t} dv\frac{\text{sgn}(f)f(v)+q_E v\bigg(e^{\chi} v A_t(v)(\mathcal{E}+q_E A_t(v))+A_v(v)f(v)\sqrt{e^{\chi(v)}v^2(\mathcal{E}+q_E A_t (v))^2+f(v)}\bigg)}{v f(v)\sqrt{e^{\chi(v)}v^2(\mathcal{E}+q_E A_t (v))^2+f(v)}}
    \label{chargedaction}
\end{equation}
We find the boundary time, in this case, is
\begin{equation}
    t_b=-\int_{0}^{v_t}\frac{dv}{\dot{v}}=-\int_{0}^{v_t}\frac{e^{\chi/2}dv}{f(v)\sqrt{1+{f(v)e^{-\chi(v)}}/(v \mathcal{E}+v q_E A_t(v))^2}}\,.
    \label{boundarytime}
\end{equation}
Now the relation in (\ref{energyturn}) modifies into,
\begin{equation}
    \mathcal{E}=-q_E A_t(v_t)+\sqrt{-\frac{f(v_t)e^{-\chi(v_t)}}{v_t^2}}
\end{equation}
This new relation between the energy and turning point includes an additional term $-q_E A_t(v_t)$ compared to the previous one in \eqref{energyturn}. This additional term provides us with more flexibility to explore the interior of the black hole. In fact, by examining the behavior of $\mathcal{E}$ as a function of $v_t$ and $q_E$, we realize that this extra term is sufficient to ensure that these geodesics could reach regions very close to the singularity. We have visualized this behavior in Figure \ref{energyvsturn}. In summary, when $q_E>0$, the spacelike geodesics explore deeper regions than their neutral counterparts, whereas for $q_E<0$, they explore less of the interior compared to the neutral ones.

We now investigate whether these charged geodesics can directly probe the singularity. To address this, we examine their behavior within the Kasner regime, where
\begin{equation}
    \mathcal{E}=-q_E A_t(v_t)+v_t^{-\frac{2p_t}{1-p_t}}
\end{equation}
In the strict limit $v_t\to\infty$, $\mathcal{E}\to -q_E \,\mu_\infty$ as $0\leq p_t<1$ and $A_t(v)\to\mu_\infty=\text{constant}$\footnote{In fact, $\mu_\infty<0$ inside the black hole.} in the black hole interior. This shows that, for finite $q_E$, energies are finite in the Kasner regime and not zero, which was the case for the uncharged correlators. If we allow larger values for $q_E$, then the maximum values for the energy will be near the singularity and will be given by $-q_E \,\mu_\infty$. This means that for $q_E\to+\infty$ we have $\mathcal{E}\to\infty$ implying that the charged geodesics are capable of reaching the singularity. In this limit, we can find the leading non-analytic contributions in (\ref{chargedaction}), which comes from the Kasner regime,\footnote{This term only corresponds to the leading behavior coming from the upper integration limit of (\ref{chargedaction}). There are further terms that come from the boundary and rest of the flow, which may or may not be analytic, and we do not include them here.}
\begin{equation}
    S_{\text{on-shell}}\sim q_E \bigg[\mu_\infty\big(\mathcal{E}+q_E\,\mu_\infty\big)^{-1+\frac{1}{p_t}}+2\tilde{\mu}_\infty \big(\mathcal{E}+q_E\,\mu_\infty\big)^{\frac{-1+p_t}{2p_t}}\bigg]\,,
\end{equation}
where $\tilde{\mu}_\infty=A_v(v\to\infty)$. Likewise, we find that $t_b$ in this limit is given by\footnote{This is the leading order approximation of the boundary time.}
\begin{equation}
    t_b=t_c+\frac{1}{\mathcal{E}+q_E\, \mu}\,,
\end{equation}
where $t_c$ is the critical time given by
\begin{equation}
    t_c=\int_0^\infty \frac{e^{\chi(v)/2}}{f(v)}dv\,.
\end{equation}
Combined, we can write
\begin{equation}
    S_{\text{on-shell}}\sim q_E \bigg[\mu_\infty\left(\frac{t_q(t_b-t_c)}{t_q+t_b-t_c}\right)^{1-\frac{1}{p_t}}+2 \tilde{\mu}_\infty\left(\frac{t_q(t_b-t_c)}{t_q+t_b-t_c}\right)^{\frac{1-p_t}{2p_t}}\bigg]\,,
\end{equation}
where we have defined
\begin{equation}
t_q\equiv\frac{1}{q_E (\mu_\infty-\mu)}\,,
\end{equation}
a timescale characterizing the RG flow. Since $p_t$ may take any value in the range $0\leq p_t<1$, this shows that one can diagnose the singularity entirely from the (non-)analytic properties of the charged correlator in the limit of infinite $q/m$.

\subsection{Extremal area surfaces: entanglement and complexity rates}

We now deal with two quantities that probe the black hole's interior, complexity and entanglement entropy. They have a clear interpretation in the dual QFT but the most interesting part for us is the fact that, via holography, they are useful probes for the emergence of spacetime \cite{Lashkari:2013koa,Faulkner:2013ica,Swingle:2014uza,Agon:2020mvu,Agon:2021tia,Pedraza:2021mkh,Pedraza:2021fgp,Pedraza:2022dqi,Carrasco:2023fcj}. In particular, their computation in the bulk depends on the region inside the horizon, encoding some of the interior geometry information. To be precise, here we calculate their late-time `velocities', which are related to their rates of change with respect to the boundary time coordinate as we clarify shortly.

Complexity is a concept originating from computer science, \emph{computational complexity}, that measures how hard it is to perform a certain task. Quantitatively, we can think of it as the number of simple steps it takes to get from an elementary initial state to a specific final one. Quantum mechanically, these states are naturally quantum states and the simple steps are unitary operators acting on them, i.e. the gates of a quantum circuit. Complexity is then the smallest size of the circuit required to get to the quantum state at hand. 

Entanglement entropy, on the other hand, must be defined with respect to a certain subregion $\pazocal{A}$ of a Cauchy slice $\Sigma$ in the background geometry of the QFT. Then, one can say that the entanglement entropy of that subregion, $\pazocal{S}_{\pazocal{A}}\,$, provides a measure of the level of entanglement the effective degrees of freedom in $\pazocal{A}$ have with the complementary subspace $\bar{\pazocal{A}}$ ($\pazocal{A} \cup \bar{\pazocal{A}} = \Sigma$).

In the context of AdS/CFT, there are proposals of geometric procedures to compute these quantities. Regarding the latter, the Ryu-Takayanagi (RT) prescription~\cite{Ryu:2006bv,Hubeny:2007xt} proposed that the entanglement entropy of the subregion $\pazocal{A}$ is dual to the area of the extremal codimension-$2$ hypersurface, $\Gamma_{\pazocal{A}}$, anchored by $\partial\pazocal{A}$ (boundary of $\pazocal{A}$) at both CFTs of the corresponding eternal AdS-black hole spacetime,
\be
    \label{eq:ch6_3_entanglement}
    \pazocal{S}_{\pazocal{A}} = \text{ext}\left [  \frac{\text{Area}(\Gamma_{\pazocal{A}})}{4G_{(d+1)}} \right ] \;.
\ee
Unfortunately, in the majority of systems, these extremal hypersurfaces do not foliate the whole interior geometry, reaching a maximum $v_c>v_+$ as the boundary time approaches infinity $t_b \rightarrow \infty$. Therefore, this quantity does not effectively probe the near-singularity geometry, though it does capture some information about the trans-IR regime.

As for complexity, the posterior CV proposal~\cite{Susskind:2014rva,Stanford:2014jda,Susskind:2014moa} is very similar to the RT prescription. In this case, the extremal hypersurface must only be anchored at the boundary time $t_b$ of the CFTs, yielding a codimension-$1$ hypersurface (instead of codimension-$2$) which is precisely the Einstein-Rosen bridge. Thus, the name CV comes from `$\text{Complexity} = \text{Volume}$' as in
\be
    \label{eq:ch6_3_complexity}
    C_V = \text{max} \left [ \frac{\text{Vol}(\pazocal{B}_{\sigma})}{l G_{(d+1)}} \right ] \;,
\ee
where $\pazocal{B}$ is the Einstein-Rosen bridge, $\sigma$ is $\sigma = \partial \pazocal{B}$ and $l$ an arbitrary length scale usually chosen to be $L$ of AdS. Despite this distinction, complexity is also limited from probing the near-singularity geometry in the same way as the entanglement entropy.\footnote{Here, we focus exclusively on the CV proposal for two key reasons: its computational similarity to the RT prescription and its relatively well-understood quantum circuit interpretation among complexity proposals \cite{Pedraza:2021mkh,Pedraza:2021fgp}. However, other complexity measures, such as `Complexity = Action' \cite{Brown:2015bva,Brown:2015lvg} and `Complexity = Anything' \cite{Belin:2021bga} could in principle probe the region near the singularity more directly \cite{Jorstad:2023kmq,Arean:2024pzo}. We leave the exploration of these alternatives for future work.} The collapse of the Einstein-Rosen bridge makes it even more difficult to access the singularity, as $v_c$ remains very close to the horizon in this scenario. We may then say that the positive $p_t$ Kasner singularity `repels the Einstein-Rosen bridge'.

It is also important to note that, in these proposals, both complexity and entanglement entropy grow linearly with the boundary time $t_b$ at late times. Consequently, it is more interesting to study the slope of their growth (velocity) as a function of the physical dimensionless ratios determining the background.

Given the ansatz~\eqref{eq:EC_metric_ansatz}, the authors of~\cite{Caceres:2022smh} computed a general formula for the volume density of codimension-$(k+1)$ extremal hypersurfaces among which are complexity ($k=0$) and entanglement entropy ($k=1$). We rewrite the formula here\footnote{This is a UV divergent quantity as we are integrating from $v=0$. However, if we want to compute \eqref{eq:ch6_3_codimensionk1} explicitly, then we must regulate it.}
\be
    \label{eq:ch6_3_codimensionk1}
    \pazocal{V}_k(\pazocal{E}) = \frac{2}{\pazocal{E}}\int_{0}^{v_m} \frac{dv}{v^{2(d-k)}} \frac{e^{-\Tilde{\chi}(v)/2}}{\sqrt{1+f(v)e^{-\Tilde{\chi}(v)}/(v^{d-k}\pazocal{E})^2}} \;,
\ee
where $v_m$ is the `turnaround' point of the surface (the maximal radius it reaches inside the black hole) given by
\be
    \label{eq:ch6_3_energy2}
    \pazocal{E}^2 = \frac{\left | f(v_m) \right | e^{-\Tilde{\chi}(v_m)}}{v_{m}^{2(d-k)}} \;,
\ee
and the boundary time is computed from
\be
    \label{eq:ch6_3_bdytime}
    t_b = - P \int_{0}^{v_m}\frac{sgn(\pazocal{E})e^{\Tilde{\chi}(v)/2}}{f(v)\sqrt{1+f(v)e^{-\Tilde{\chi}(v)}/(v^{d-k}\pazocal{E})^2}} \;.
\ee
Now, differentiating~\eqref{eq:ch6_3_codimensionk1} and~\eqref{eq:ch6_3_bdytime}, and using~\eqref{eq:ch6_3_energy2} we arrive at
\be
    \label{eq:ch6_3_velocityk}
    \frac{d\pazocal{V}_k}{dt_b}\Biggr\rvert_{v=v_c} = \sqrt{2} \, \frac{\sqrt{\left | f(v) \right |}e^{-\Tilde{\chi}(v)/2}}{v^{d-k}}\Biggr\rvert_{v=v_c} \;,
\ee
which, with the appropriate normalization as given in~\cite{Frenkel:2020ysx}, results in 
\be
    \label{eq:ch6_3_velocityk_def}
    \frac{d\pazocal{V}_k}{dt_b}\Biggr\rvert_{v=v_c} = v_{+}^{d-k} \, \frac{\sqrt{\left | f(v) \right |}e^{-\Tilde{\chi}(v)/2}}{v^{d-k}}\Biggr\rvert_{v=v_c} \;.
\ee
\begin{figure}[t]
    \centering
    \includegraphics[scale=0.53]{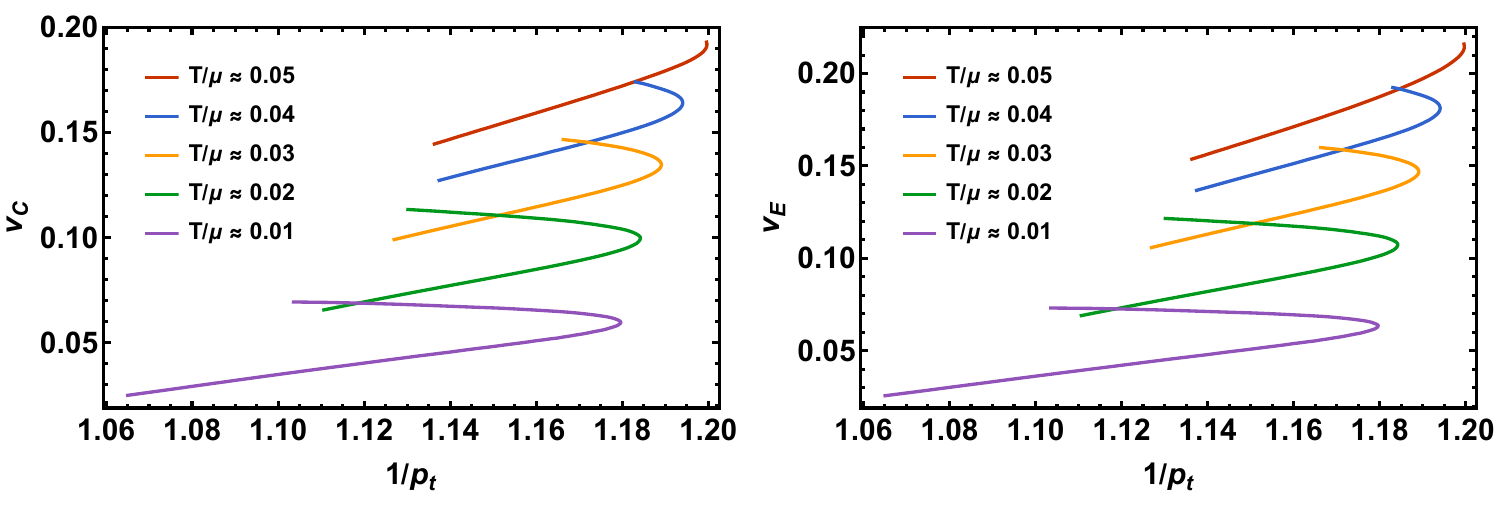}
    \caption{Complexity (left) and entanglement (right) velocities for the deformed EC system computed from~\eqref{eq:ch6_3_velocityk_def} ($k=0$ and $k=1$ respectively) as a function of the inverse of the Kasner exponent $p_t$, for several fixed values of $T/\mu$. The cloud input parameters are: $\hat{\beta}=10$, $\hat{m}=0.55$ and $\hat{q} \simeq 2.23\,$.}
    \label{fig:ch6_3_velocities_phi}
\end{figure}
\begin{figure}[t]
    \centering
    \includegraphics[scale=0.53]{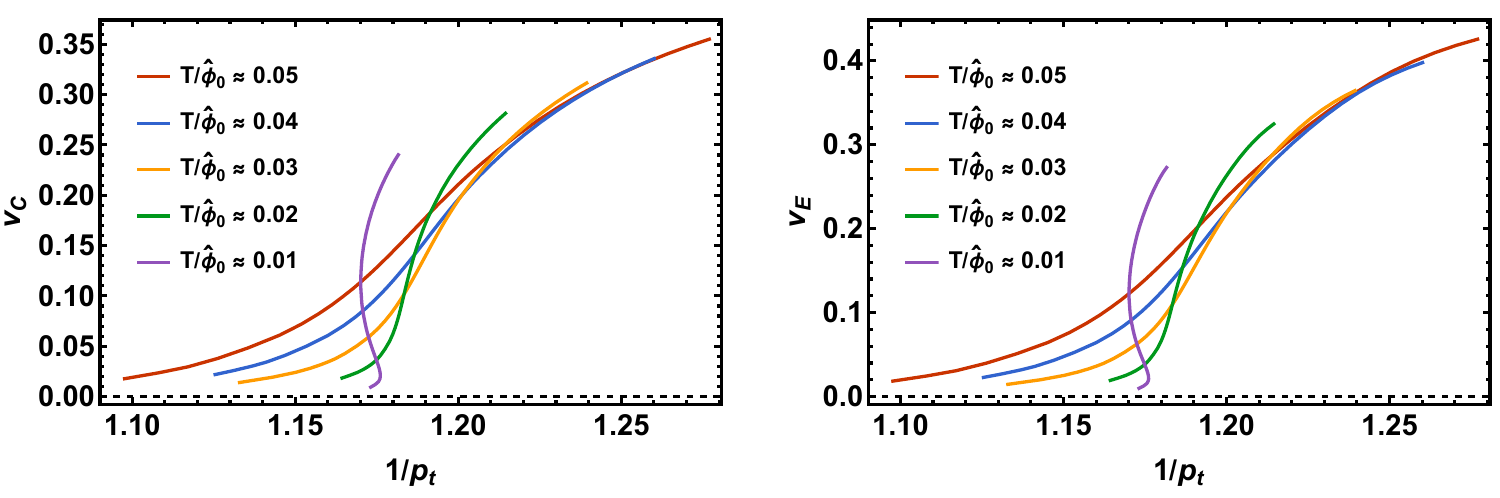}
    \caption{Complexity (left) and entanglement (right) velocities for the deformed EC system computed from~\eqref{eq:ch6_3_velocityk_def} ($k=0$ and $k=1$ respectively) as a function of the inverse of the Kasner exponent $p_t$, for several fixed values of $T/\hat{\phi}_0$. The cloud input parameters are: $\hat{\beta}=10$, $\hat{m}=0.55$ and $\hat{q} \simeq 2.23\,$.}
    \label{fig:ch6_3_velocities_mu}
\end{figure}

The results for the complexity and entanglement velocities of the deformed electron cloud system are shown in Figures~\ref{fig:ch6_3_velocities_phi} and~\ref{fig:ch6_3_velocities_mu}. Although these velocities do not reach the near-singularity region, we plot them as a function of the Kasner exponent $p_t$ to highlight that they contain information intrinsic to the black hole's interior.
According to Figure~\ref{fig:ch5_1_kasner_pt}, we conclude that these velocities generally increase with the temperature, while their dependence on $p_t$ is multivalued. This behavior can be linked from the multivaluedness of $p_t$ as a function of $\hat{\phi}_0/T$, which can be inferred from Figure~\ref{fig:ch5_1_kasner_pt}. Additionally, the entanglement velocity takes slightly higher values than its partner for the same set of backgrounds. 

\subsection{Butterfly velocity and operator spreading}
In quantum many-body dynamics, the butterfly velocity~\cite{Shenker:2013pqa,Roberts:2014isa} quantifies the propagation speed of the influence caused by local perturbations to the system. It represents a measure of how chaotic the system is, hence the name `butterfly' (alluding to the butterfly effect).

These local perturbations are realized by insertions of hermitian operators, such that this quantity can be calculated through the effective size of the commutator between them as $\langle \left [ 
W(x,t),\,V(0,0) \right ]^2\rangle_{\beta} \sim e^{\lambda_L(t-x/v_B)}\,$, where $\langle \cdot \rangle_{\beta}$ is the thermal average ($\beta = 1/T$), $\lambda_L$ is known as the Lyapunov exponent and $v_B$ is the butterfly velocity. 

The Lyapunov exponent, measuring the influence of global perturbations through time, has an upper bound in general quantum systems, $\lambda_L \leq 2\pi /\beta$, which is always saturated when they are holographic (admitting an Einstein gravity dual)~\cite{Maldacena:2015waa}. Similarly, the butterfly velocity was proven~\cite{Mezei:2016zxg} to be bounded by $v_B \leq \sqrt{\frac{d}{2(d-1)}}$ in Einstein gravity AdS-black hole systems. However, various works~\cite{Giataganas:2017koz,Fischler:2018kwt,Gursoy:2020kjd,Eccles:2021zum} have shown that violations of this bound might happen in holographic systems that break relativistic invariance along the RG flow. We are thus interested in studying whether or not this upper limit is adhered to in DEC systems, given that it is Lifshitz invariant in the IR, hence non-relativistic.

In the context of AdS/CFT, the local insertions of operators correspond to shooting shock waves into the black hole in the bulk. The butterfly velocity can then be computed by studying the backreaction from this shock wave geometry, as explained in~\cite{Roberts:2016wdl}. In this work we compute the butterfly velocity of the deformed electron cloud system via a different method from the shock wave geometry, first presented in~\cite{Mezei:2016wfz}. The authors of~\cite{DiNunno:2021eyf} followed this method to derive the butterfly velocity
\be
    \label{eq:ch6_2_butterfly}
    v_B = \sqrt{\frac{g_{tt}'(v_+)}{2g_{ii}'(v_+)}} \;,
\ee
solely using a general formula for metrics of the form
\be
    \label{eq:ch6_2_metric_vb}
    ds^2 = -g_{tt}(v)dt^2 + g_{vv}(v)dv^2 +g_{ii}(v)d\Vec{x}^{\,2} \;.
\ee

\begin{figure}[t]
    \centering
    \includegraphics[scale=0.55]{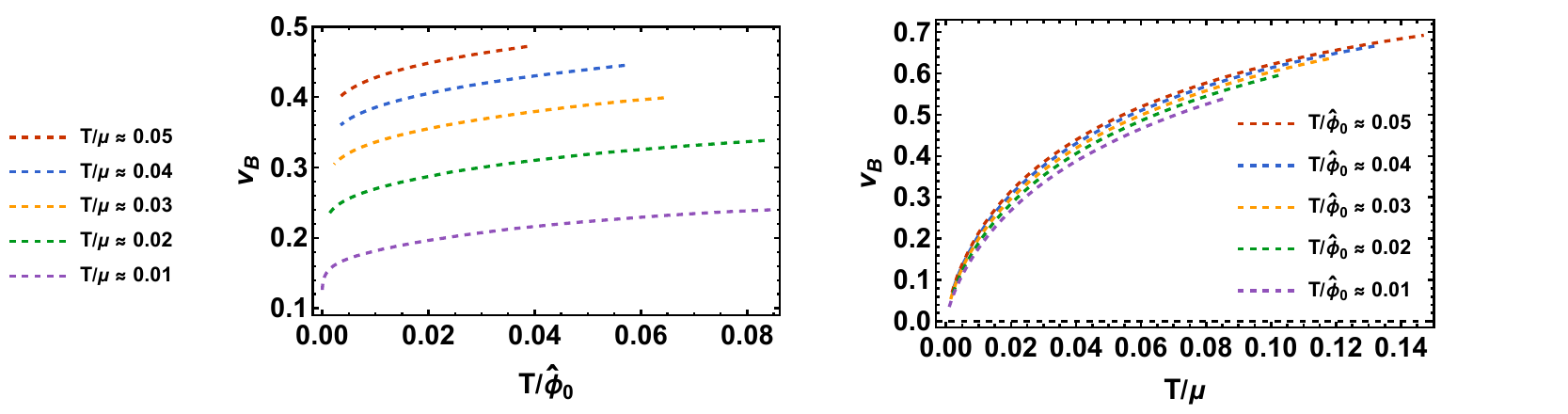}
    \caption{Butterfly velocity as a function of $T/\hat{\phi}_0$ for several fixed values of $T/\mu$ (left) and as a function of $T/\mu$ for several fixed values of $T/\hat{\phi}_0$ (right), in the deformed EC system. The cloud input parameters are: $\hat{\beta}=10$, $\hat{m}=0.55$ and $\hat{q} \simeq 2.23\,$.}
    \label{fig:ch6_2_vB}
\end{figure}

We present the results over many backgrounds using~\eqref{eq:ch6_2_butterfly} in Figure~\ref{fig:ch6_2_vB}. We note that this observable is completely determined in terms of near-horizon data, so we have refrained from plotting its behavior with respect to $p_t$. As the temperature takes higher values, both with respect to $\hat{\phi}_0$ and $\mu$, the butterfly velocity monotonically increases. This hints us that the bound of $v_B \leq \sqrt{3/4} \simeq 0.87$ (for $d=3$) holds despite the Lifshitz IR fixed point, since at higher temperatures the EC disappears and the system becomes relativistic at all scales.
Further, similarly to~\cite{DiNunno:2021eyf}, it also seems that $v_B$ approaches a constant non-zero value as $T/\hat{\phi}_0 \rightarrow 0$, which indicates that its behavior in this limit may be used to characterize the IR fixed point, i.e., to extract $z$. 

\begin{figure}[t]
    \centering
    \includegraphics[scale=0.57]{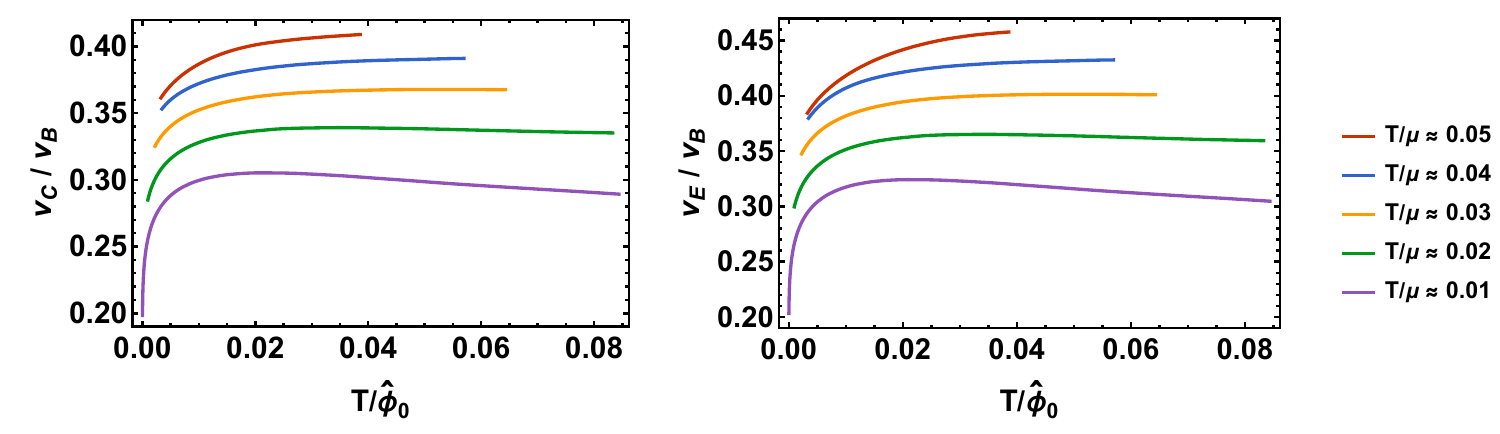}
    \caption{Ratios of the complexity (left) and entanglement (right) velocities over the butterfly velocity for the deformed EC system as a function of $T/\hat{\phi}_0\,$, for several fixed values of $T/\mu\,$. The cloud input parameters are: $\hat{\beta}=10$, $\hat{m}=0.55$ and $\hat{q} \simeq 2.23\,$.}
    \label{fig:ch6_3_velocities_ratios_phi}
\end{figure}

\begin{figure}[t]
    \centering
    \includegraphics[scale=0.52]{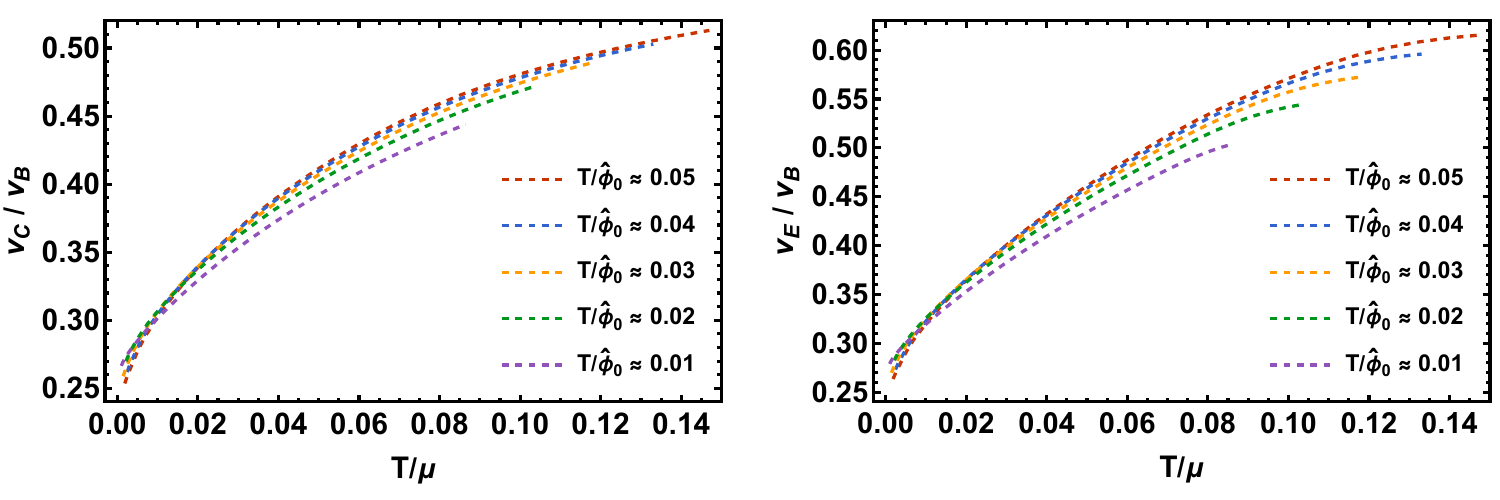}
    \caption{Ratios of the complexity (left) and entanglement (right) velocities over the butterfly velocity for the deformed EC system as a function of $T/\mu\,$, for several fixed values of $T/\hat{\phi}_0\,$. The cloud input parameters are: $\hat{\beta}=10$, $\hat{m}=0.55$ and $\hat{q} \simeq 2.23\,$.}
    \label{fig:ch6_3_velocities_ratios_mu}
\end{figure}

Finally, we perform a cross-check of our numerics in Figures~\ref{fig:ch6_3_velocities_ratios_phi} and~\ref{fig:ch6_3_velocities_ratios_mu}. Here, we confirm that the complexity and entanglement velocities are smaller than the butterfly velocity. This must always be the case as the butterfly velocity, per its definition, characterizes an effective light cone in the field theory, bounding the spreading of information.

\section{Discussion\label{sec:discussion}}

In this paper, we investigated the gravitational dual of a fermionic field theory at finite temperature and charge density in $(2+1)$-dimensions. Our focus was on a $(3+1)$-dimensional Einstein-Maxwell system coupled to a free fermion fluid, often described as an electron cloud. We introduced a relevant scalar operator to this system, triggering a renormalization group flow that significantly altered the geometry within the bulk black hole. Our findings provide valuable insights into the properties and dynamics of holographic metals, particularly highlighting the impact of deformations on their internal structure and observables.

One of the key findings of our work is the destruction of the inner (Cauchy) horizon due to the deformation, aligning with existing no-inner-horizon theorems for hairy black holes \cite{Hartnoll:2020rwq,Cai:2020wrp,An:2021plu,Arean:2024pzo}. This deformation results in a near-singularity metric characterized by a Kasner cosmology, with its specifics depending on the strength of the deformation. Despite this variation, the Kasner exponent $p_t$ remains positive, consistently signaling the collapse of the Einstein-Rosen bridge. This outcome stands in stark contrast to the undeformed scenario, where the interior geometry resembles that of an AdS-RN black hole, featuring a timelike singularity.\footnote{As a result of Fermi seasickness, the electron cloud system expels some of its charge at low temperatures. However, we do not observe the characteristic Josephson oscillations typically found in holographic superconducting models. This can be attributed to the fact that the electron cloud is confined to the black hole exterior, while the scalar field, responsible for the backreaction in the black hole interior, is neutral \cite{Mansoori:2021wxf}. It is tempting to speculate whether deformations of the EC system by a charged operator \cite{Liu:2013yaa} could result in interior geometries resembling those of a holographic superconductor; we leave this exploration for future work.} Additionally, we observed that the deformed electron cloud exhibits Lifshitz scaling symmetry in the IR region at low temperatures, characterized by a dynamical critical exponent $z$ that depends on the strength of the deformation. This exponent breaks Lorentz invariance by enforcing distinct scaling behaviors for time and spatial coordinates. Such non-relativistic scaling symmetry is characteristic of strange metal behavior near quantum critical points, highlighting the complex nature of the dual field theory.

The analysis of the thermal $a$-function and its derivative provided deeper insights into the RG flow and the fixed points of the theory. The monotonic decrease of the $a$-function along the holographic radial direction reflects a systematic reduction in degrees of freedom from the UV to the IR. Moreover, by pinpointing where the $a$-function becomes stationary, we discovered that the system exhibits both a Lifshitz IR fixed point and an additional Kasner trans-IR fixed point ---absent in neutral RG flows--- making this a novel discovery in our context. This additional fixed point, characterized by the value of $p_t$, represents a state where the degrees of freedom effectively freeze out, underscoring the significant impact of the scalar deformation on the system's dynamics.

We further explored various probes of the RG flows and black hole interiors. Analyzing spacelike geodesics and their lengths in the near-singularity region offered insights into the spacelike correlations in the dual field theory. By calculating two-point correlation functions for neutral and charged operators using their geodesic approximations, we found that charged operators can penetrate deeper into the black hole interior than their neutral counterparts. This highlights the influence of the Maxwell field and scalar deformation on both geodesic length and the resulting correlation functions. Further, we discovered that charged geodesics with $q/m\to\infty$ can reach the singularity despite the positive-$p_t$ Kasner exponent. In particular, this implies that one can diagnose the singularity by looking at non-analyticities in charged correlators. Using holographic prescriptions for complexity and entanglement entropy, we further investigated how the deformation affects the interior geometry. Our results showed that both observables increase linearly with boundary time at late times, with their growth rates providing insights into the interior structure. Notably, the velocities of entanglement and complexity display multivalued behavior with respect to $p_t$, reflecting the complex influence of the interior geometry. Finally, we computed the butterfly velocity, a key indicator of operator spreading in the dual theory. Although this observable is entirely determined by horizon data, we demonstrated that it provides a reliable upper bound for the rates of complexity and entanglement growth, both of which are dependent on the interior geometry.

In summary, our investigation into the electron cloud under relevant deformations provides a thorough framework for understanding the dynamics of holographic metals. The destruction of the inner horizon, the emergence of Lifshitz and Kasner fixed points, and the detailed analysis of RG flows and their observables collectively present a rich and nuanced picture of the critical phenomena and transport properties of these systems. Our findings significantly advance the understanding of strongly correlated electronic systems and their holographic duals, paving the way for future research, particularly in exploring new deformations and their effects on gravitational duals of condensed matter systems.

\noindent \noindent\section*{Acknowledgments}
We are grateful to Daniel Areán, Elena Cáceres, Hyun-Sik Jeong, Miguel Montero, Ángel Murcia, Gerben Oling and Le-Chen Qu for useful discussions and correspondence. JC is
supported by the Royal Society Research Grant RF/ERE/210267. AKP and JFP are supported by the ‘Atracci\'on de Talento’ program (Comunidad de Madrid) grant 2020-T1/TIC-20495, by the Spanish Research Agency via grants CEX2020-001007-S and PID2021-123017NB-I00, funded by MCIN/AEI/10.13039/501100011033, and by ERDF A way of making Europe.

\appendix

\bibliographystyle{JHEP}
\bibliography{manuscript.bib}

\end{document}